\documentclass[
  aip, 
  apl, 
  amsmath, 
  amssymb, 
  reprint]{revtex4-2}

\usepackage[utf8]{inputenc} 
\usepackage[T1]{fontenc} 
\usepackage{mathptmx} 
\usepackage{graphicx}
\usepackage{braket} 
\usepackage{hyperref} 
\hypersetup{
  colorlinks=true,
  citecolor=blue,
  linkcolor=blue,
  urlcolor=blue}

\newcommand{\supplementary}{\href{https://www.scitation.org}{supplementary material}}

\newcommand{\Figureautorefname}{Figure}

\newcommand{\figref}[2][]{%
  \hyperref[#2]{\Figureautorefname~\ref*{#2}#1}%
}

\makeatletter
\renewcommand{\fnum@figure}{\textbf{FIG. \thefigure}}
\makeatother

\begin{document}

\title{Pursuing high-fidelity control of spin qubits in natural Si/SiGe quantum dot} 

\author{Ning Wang}
\thanks{These authors contributed equally to this work}
\affiliation{\mbox{CAS Key Laboratory of Quantum Information, University of Science and Technology of China, Hefei, Anhui 230026, China}}
\affiliation{\mbox{CAS Center for Excellence in Quantum Information and Quantum Physics, University of Science and Technology of China,} \\Hefei, Anhui 230026, China}

\author{Shao-Min Wang}
\thanks{These authors contributed equally to this work}
\affiliation{\mbox{CAS Key Laboratory of Quantum Information, University of Science and Technology of China, Hefei, Anhui 230026, China}}
\affiliation{\mbox{CAS Center for Excellence in Quantum Information and Quantum Physics, University of Science and Technology of China,} \\Hefei, Anhui 230026, China}

\author{Run-Ze Zhang}
\affiliation{\mbox{CAS Key Laboratory of Quantum Information, University of Science and Technology of China, Hefei, Anhui 230026, China}}
\affiliation{\mbox{CAS Center for Excellence in Quantum Information and Quantum Physics, University of Science and Technology of China,} \\Hefei, Anhui 230026, China}

\author{Jia-Min Kang}
\affiliation{\mbox{CAS Key Laboratory of Quantum Information, University of Science and Technology of China, Hefei, Anhui 230026, China}}
\affiliation{\mbox{CAS Center for Excellence in Quantum Information and Quantum Physics, University of Science and Technology of China,} \\Hefei, Anhui 230026, China}

\author{Wen-Long Lu}
\affiliation{\mbox{CAS Key Laboratory of Quantum Information, University of Science and Technology of China, Hefei, Anhui 230026, China}}
\affiliation{\mbox{CAS Center for Excellence in Quantum Information and Quantum Physics, University of Science and Technology of China,} \\Hefei, Anhui 230026, China}

\author{Hai-Ou Li}
\affiliation{\mbox{CAS Key Laboratory of Quantum Information, University of Science and Technology of China, Hefei, Anhui 230026, China}}
\affiliation{\mbox{CAS Center for Excellence in Quantum Information and Quantum Physics, University of Science and Technology of China,} \\Hefei, Anhui 230026, China}
\affiliation{\mbox{Hefei National Laboratory, University of Science and Technology of China, Hefei 230088, China}}

\author{Gang Cao}
\affiliation{\mbox{CAS Key Laboratory of Quantum Information, University of Science and Technology of China, Hefei, Anhui 230026, China}}
\affiliation{\mbox{CAS Center for Excellence in Quantum Information and Quantum Physics, University of Science and Technology of China,} \\Hefei, Anhui 230026, China}
\affiliation{\mbox{Hefei National Laboratory, University of Science and Technology of China, Hefei 230088, China}}

\author{Bao-Chuan Wang}
\thanks{Corresponding author: \href{mailto:bchwang@ustc.edu.cn}{bchwang@ustc.edu.cn}}
\affiliation{\mbox{CAS Key Laboratory of Quantum Information, University of Science and Technology of China, Hefei, Anhui 230026, China}}
\affiliation{\mbox{CAS Center for Excellence in Quantum Information and Quantum Physics, University of Science and Technology of China,} \\Hefei, Anhui 230026, China}

\author{Guo-Ping Guo}
\affiliation{\mbox{CAS Key Laboratory of Quantum Information, University of Science and Technology of China, Hefei, Anhui 230026, China}}
\affiliation{\mbox{CAS Center for Excellence in Quantum Information and Quantum Physics, University of Science and Technology of China,} \\Hefei, Anhui 230026, China}
\affiliation{\mbox{Hefei National Laboratory, University of Science and Technology of China, Hefei 230088, China}}
\affiliation{\mbox{Origin Quantum Computing Company Limited, Hefei, Anhui 230026, China}}

\date{\today}

\begin{abstract}
  Electron spin qubits in silicon are a promising platform for fault-tolerant quantum computing. Low-frequency noise, including nuclear spin fluctuations and charge noise, is a primary factor limiting gate fidelities. Suppressing this noise is crucial for high-fidelity qubit operations. Here, we report on a two-qubit quantum device in natural silicon with universal qubit control, designed to investigate the upper limits of gate fidelities in a non-purified Si/SiGe quantum dot device. By employing advanced device structures, qubit manipulation techniques, and optimization methods, we have achieved single-qubit gate fidelities exceeding 99\% and a two-qubit Controlled-Z (CZ) gate fidelity of 91\%. Decoupled CZ gates are used to prepare Bell states with a fidelity of 91\%, typically exceeding previously reported values in natural silicon devices. These results underscore that even natural silicon has the potential to achieve high-fidelity gate operations, particularly with further optimization methods to suppress low-frequency noise.

\end{abstract}

\maketitle 

Quantum computing aims to solve complex problems that are intractable for classical computers due to their exponential complexity with increasing dimensions. Practical quantum computers require quantum error correction to protect qubits from losing coherence, necessitating a large number of qubits operating with high fidelities.\cite{RN10693,RN11492} Spin qubits in semiconductor quantum dots are advantageous among solid-state systems due to their relatively long coherence times,\cite{RN1302} nanoscale feature sizes,\cite{RN7360,RN11424} and compatibility with semiconductor industry processes,\cite{RN11647,RN10696,RN11981} offering significant scalability potential. A crucial step is to achieve high-fidelity qubit gate operations, which requires a comprehensive understanding of the mechanisms that limit gate fidelities.

For spin qubits in Si/SiGe quantum dots, one primary factor limiting gate fidelities is nuclear spins, which couple with electron spins through hyperfine interactions, thus reducing electron spin dephasing times.\cite{RN4285} Early work on natural silicon has achieved high-fidelity single-qubit operations\cite{RN4323, RN1441, RN1564} and demonstrated two-qubit gates, albeit with low fidelity\cite{RN1419, RN1413, RN9045}. Recent progress with isotopic purification to reduce non-zero nuclear spins and the optimization of qubit control methods has resulted in two-qubit gate fidelities exceeding 99\%\cite{RN10510, RN10511, RN10653} and universal control of up to six qubits.\cite{RN11493} However, the improvement in dephasing times from isotope purification has not been as significant as anticipated, typically remaining at only a few microseconds, likely limited by charge noise mediated by micromagnets rather than nuclear spin noise. This motivates further exploration of the potential for achieving high-fidelity gate operations in natural silicon, focusing on addressing charge and nuclear noises.

Here, we report on a two-qubit quantum device fabricated on a natural Si/SiGe heterostructure, designed to explore the upper limits of gate fidelity in non-purified silicon. To achieve high-fidelity qubit control, including single- and two-qubit gates, state-of-the-art procedures are implemented to characterize and calibrate qubit parameters such as Rabi frequency, exchange interaction strength, and two-qubit operation points. The gate performance is evaluated using randomized benchmarking protocols. Additionally, decoupled CZ gates are employed to suppress low-frequency noise, enabling the preparation of Bell states with high fidelity. Finally, we discuss the limitations of our device and propose further improvements to achieve high-fidelity qubit gate operations even in non-purified silicon. 

\begin{figure*}
  \centering
  \includegraphics{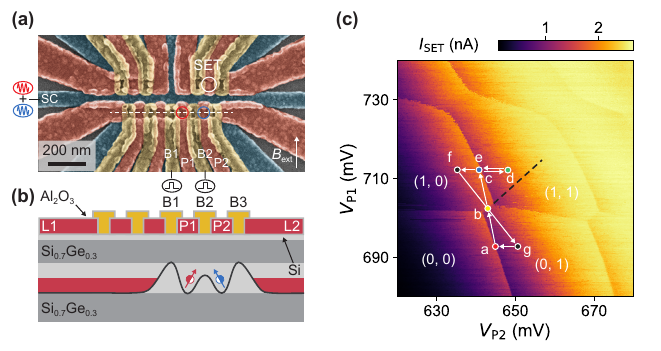}
  \caption{\label{fig1}
  (a) False-colored SEM image of the device with a schematic cross-section of the DQD (along the dashed line). Two spin qubits are formed under gates P1 and P2, with barrier gates B1 and B2 controlling the exchange coupling between them. 
  (b) Charge stability diagram with pulse sequences for initialization, operation, readout, and emptying. The black dashed line indicates the position where the dot chemical potential detuning is zero.}
\end{figure*}

\figref{fig1}(a) shows a false-colored scanning electron microscope (SEM) image of a four-dot linear array alongside a schematic cross-section, identical to the device used in this experiment. The device is fabricated on an undoped natural Si/SiGe heterostructure using an overlapping gate architecture with aluminum gates, similar to our previous work.\cite{RN11424} More details about the device and measurement setup are shown in \supplementary. A double quantum dot (DQD) is formed under gates P1 and P2, with gate B2 controlling the tunnel coupling between them. The gates on the left side are set at the same voltage as gate L1, serving as electron reservoirs. Gate B1 (B3) controls the tunneling rate between dot 1 (2) and its respective reservoir. The spin states of the single electron are used to encode spin qubits, and an external magnetic field $B_\text{ext}=500 \ \text{mT}$ is applied parallel to the quantum well to induce Zeeman splitting. An integrated Co micromagnet provides both longitudinal and transverse magnetic field gradients for qubit addressing and control, respectively.\cite{RN1760,RN1578} The larger Zeeman energy compared to thermal energy ($T_\text{e} \approx 150 \ \text{mK}$) allows for the initialization and readout of spin qubits via energy-selective tunneling through nearby electron reservoirs. 

\figref{fig1}(b) displays the charge-stability diagram for the DQD by measuring the current of the charge sensor dot while sweeping the plunger gate voltages $V_\text{P1}$ and $V_\text{P2}$. The DQD is configured in the $(N_1,N_2 )=(1,1)$ charge state, where $N_i$ denotes the electron number in each dot. The pulse sequences for qubit initialization, control, and readout are illustrated on the diagram. Starting at point a, the gate voltages are pulsed to point b and then to point c, initializing the two qubits in the $\ket{\downarrow \downarrow}$ state. The qubits are then pulsed to point d, deep in the $(1,1)$ charge state, where single- and two-qubit gate operations are performed. After the operations, Q2 is first read out at point e and then emptied at point f. Subsequently, Q1 is quickly shuttled from the left dot to the right dot for final readout at point a. In this scheme, Q2 can be directly initialized and readout via the right reservoir, while Q1 requires an additional shuttling process for readout.\cite{RN9241, RN12024, RN12030}

\begin{figure*}
  \includegraphics{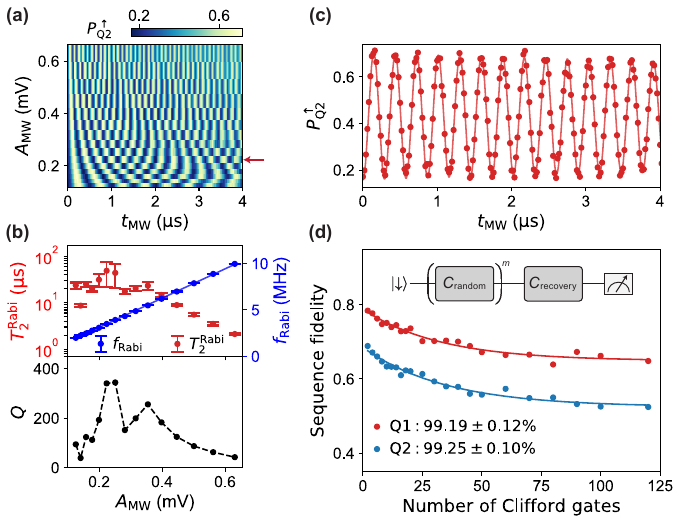}
  \caption{\label{fig2}Single-qubit logic gates. 
  (a) Rabi oscillations of Q2 as functions of microwave amplitude $A_\text{MW}$ and burst duration $t_\text{MW}$. 
  (b) Upper: $T_2^{\text{Rabi}}$ and $f_{\text{Rabi}}$ versus $A_\text{MW}$, obtained by fitting the Rabi oscillation curve with the function $P^\uparrow (t_{\text{MW}}) = A \exp(-t_{\text{MW}} / T_2^{\text{Rabi}}) \sin (2\pi f_{\text{Rabi}} t_{\text{MW}} + \phi) + B$, where $A$, $B$, $\phi$, $T_2^{\text{Rabi}}$, and $f_{\text{Rabi}}$ are fitting parameters.
  Lower: Q-factor versus $A_\text{MW}$, calculated by $Q = 2 T_2^{\text{Rabi}} f_{\text{Rabi}}$.
  (c) Rabi oscillation near the maximum $Q$ value [red arrow in (a)] with $A_{\text{MW}} \approx 0.22$ mV, $f_{\text{Rabi}} = 3.45$ MHz, and $T_2^{\text{Rabi}} = 27.6$ $\mu$s. 
  (d) Single-qubit gate fidelities measured by SRB for the two qubits. Each data point corresponds to the average of 15 random sequences, each with 30 repetitions. The data for Q2 is shifted down by 0.1 for clarity.}
\end{figure*}

Single electron spin rotation is achieved via electric-dipole spin resonance (EDSR).\cite{RN4323, RN1760} The Zeeman-energy difference between the qubits, approximately 42 MHz ($f_{\text{Q1}} = 19.235$ GHz and $f_{\text{Q2}} = 19.277$ GHz), is sufficiently larger than the exchange coupling ($<10$ MHz) and Rabi frequencies ($\sim 3$ MHz), enabling addressable and high-fidelity qubit control. Single-qubit $X\ (Y)$ gates are implemented by pulsing microwave signals at resonant frequencies, with the microwave phase determining the rotation axis. $Z$ gates are performed virtually by modifying the microwave phases. Here, an $X\ (Y)$ gate denotes a $\pi/2$ rotation around the $\hat{x}\ (\hat{y})$ axis, while a $Z$ gate denotes a $\pi/2$ rotation around the $\hat{z}$ axis. The dephasing times for Q1 (Q2) are $T_2^* = 0.9\ (0.7) \ \mathrm{\mu s}$ measured by Ramsey measurements and can be extended to $T_2^{\text{Hahn}} = 17.7\ (10.9) \ \mathrm{\mu s}$ with Hahn echo (see \supplementary).

To achieve high-fidelity single-qubit gates, we utilize a method of maximizing the quality factor (Q-factor) of Rabi oscillation to balance control speed and coherence time.\cite{RN1564,RN1540} The Q-factor is defined as $Q = T_2^{\text{Rabi}}/{T_\pi}$, where $T_\pi$ is $\pi$ rotation duration, and $T_2^{\text{Rabi}}$ is the Rabi decay time. Rabi oscillations are measured at different microwave power levels, with the results shown in \autoref{fig2}(a). The extracted Rabi frequency $f_\text{Rabi}$ and $T_2^{\text{Rabi}}$ are shown in \autoref{fig2}(b). $f_\text{Rabi}$ exhibits a linear dependence on the microwave amplitude. At low power levels, the $Q$ value is small due to the long driving time. Conversely, at high power levels, despite the increased driving speed, the significant decrease in $T_2^{\text{Rabi}}$ results in a low $Q$ value, likely due to microwave heating effects. High-quality Rabi oscillations are observed at $A_\text{MW} \approx 0.22 \text{mV}$, with a $Q$ value exceeding 300 and an $f_\text{Rabi}$ of 3.5 MHz [\autoref{fig2}(c)].

We characterize the gate fidelities using Clifford-based randomized benchmarking (RB).\cite{RN2054} Following the standard RB (SRB) sequence, we measure the spin-up probability of each qubit as a function of the number of Clifford gates ($m$), as shown in \autoref{fig2}(d). The decay is fitted by $P(m) = A p_{\text{ref}}^m + B$, where \(p_{\text{ref}}\) is the depolarizing parameter, and parameters $A$ and $B$ absorb the SPAM errors. The average fidelity per Clifford gate is determined as $F_{1 \text{Q}}^{\text{c}}=\left(1+p_{\text{ref}}\right) / 2$. Given that a Clifford gate typically includes 1.875 single-qubit rotations, we can obtain the average single-qubit gate fidelities for Q1 and Q2 to be 99.19(12)\% and 99.25(10)\%, respectively, both exceeding the threshold required for fault-tolerant quantum computing based on surface codes.\cite{RN11492}

Additionally, interleaved randomized benchmarking (IRB) is implemented to assess the fidelity of specific single-qubit gates.\cite{RN2042} Following the same fitting procedure as in SRB but substituting \(p_{\text{ref}}\) with \(p_{\text{gate}}\), the interleaved single-qubit gate fidelity is determined by $F_{\text{1Q}}^{\text{gate}} = (1 + p_{\text{gate}} / p_{\text{ref}} ) / 2$. The fidelities of $X$ and $Y$ gates are 99.67(35)\% and 99.87(40)\% for Q1, and 99.43(45)\% and 99.85(31)\% for Q2, respectively (see \supplementary). These results, consistent with SRB values, are comparable to those reported in isotope-purified silicon.\cite{RN9045,RN10510,RN10511,RN10653,RN11493}

\begin{figure}
  \includegraphics{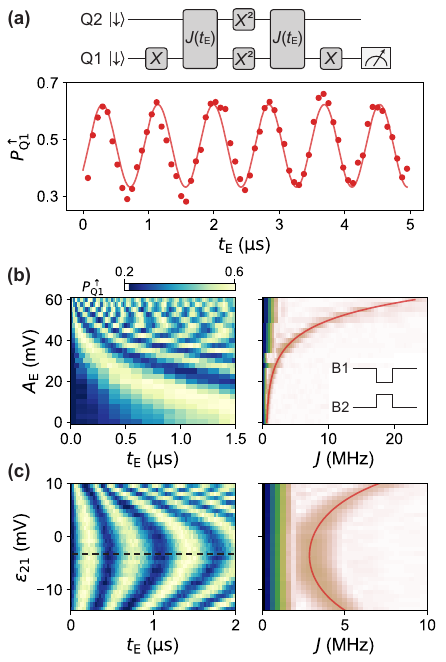}
  \caption{\label{fig3}Two-qubit parameter optimization. 
  (a) Exchange oscillations measured using a Hahn echo-type sequence with the oscillation frequency $J/2$. Q2 is the control qubit, and Q1 is the target qubit. (b) Exchange oscillation dependence on the barrier gate pulse amplitude (left) and the fast Fourier transform (FFT) results (right). The red line is an exponential fit. Inset: voltage pulses applied to barrier gates (B1 and B2) with rise and fall times of approximately 2 ns. (c) Similar to (b) but as a function of the detuning of dot chemical potentials. The black dashed line indicates the symmetrical operating point for two-qubit control, determined from the fitting of the FFT results. The red line is fitted by $J = {2t_\text{c}^2 U}/[U^2 - (\varepsilon - \varepsilon_0)^2] + J_0$, where the tunnel coupling $t_\text{c}$, detuning offset $\varepsilon_0$, and exchange offset $J_0$ are free fitting parameters.\cite{RN6050} The average charge energy $U$ is approximately 5 mV extracted from the charge transition lines in \autoref{fig1}(b).}
\end{figure}

The native two-qubit gate in our device is the controlled phase (CPhase) gate. This is achieved by applying a fast voltage pulse on the barrier gates to activate the exchange interaction ($J$) for a specific duration. The evolution of $J$ accumulates a spin-dependent phase $\phi_{1(2)} \approx {J(t)t_{\text{E}}}/{2\hbar}$
on each qubit, where $t_\text{E}$ is the evolution time. This can be described by the unitary evolution $U_{\text{CPhase}} = \text{diag}(1, e^{i\phi_2}, e^{i\phi_1}, 1)$ in the basis of $\ket{\uparrow \uparrow}, \ket{\uparrow \downarrow }, \ket{\downarrow \uparrow}$, and $\ket{\downarrow \downarrow}$. By setting $t_{\text{E}} \approx {\pi \hbar}/{J}$ to ensure $\phi_1 + \phi_2 = \pi$ and incorporating single-qubit $Z$ gates $Z_{1(2)} (-\pi/2)$ to remove the overall phases, a CZ gate $U_{\text{CZ}} = \text{diag}(1, 1, 1, -1)$ can be achieved.\cite{RN1481} The key to realizing the CZ gate lies in the precise control of the exchange interaction and the calibration of the single-qubit phases. 

To visually extract the strength of $J$, we employ a decoupled Hahn echo-type sequence [upper panel in \autoref{fig3}(a)] instead of a standard Ramsey sequence. This method avoids the impact of the unwanted frequency shift on each qubit caused by the electron wavefunction displacement within the gradient magnetic field.\cite{RN1419} The lower panel in \autoref{fig3}(a) shows the measured spin-up probability of Q1 as a function of the exchange pulse duration $t_\text{E}$, which oscillates with $J/2$. This method is also used to measure the residual coupling strength, which is approximately 150 kHz.

A large $J$ is necessary to leverage the limited coherence time. This is achieved by simultaneously applying a positive pulse to gate B2 and a negative pulse to gate B1 [inset of \autoref{fig3}(b)], effectively pushing Q1 closer to Q2 and enhancing the exchange coupling strength. The exchange oscillations are measured by stepwise increasing the exchange pulse amplitude $t_\text{E}$ and then analyzed using a fast Fourier transform (FFT), as shown in \autoref{fig3}(b). The extracted $J$ exhibits an exponential dependence on $t_\text{E}$, with a tunable range from several hundred kHz to over 20 MHz. 

Furthermore, we identify the optimal operating point (sweet spot) for two-qubit control, where $J$ is first-order insensitive to charge noise.\cite{RN6050, RN9199}
The left panel in \autoref{fig3}(c) shows the measured exchange oscillations as a function of the detuning of dot chemical potentials, $\varepsilon = \alpha (V_\text{P2}-V_\text{P1})$, where $\alpha \approx 0.12$ is the lever arm extracted from the magnetotransport experiments. We fit the FFT results to determine the symmetrical operating point. However, the sweet spot deviates from the typical position, i.e., along the extended line of the interdot charge transition, and is slightly shifted to the upper left, as shown in \autoref{fig1}(b). This deviation can be attributed to the crosstalk of barrier gates, causing an asymmetric shift between the two dot chemical potentials. Despite this, virtual gate technology can effectively mitigate the crosstalk, allowing qubits to be operated at the symmetric operation point.\cite{RN10510, RN10511, RN10653, RN11493, RN9013, RN11423}

\begin{figure*}
  \includegraphics{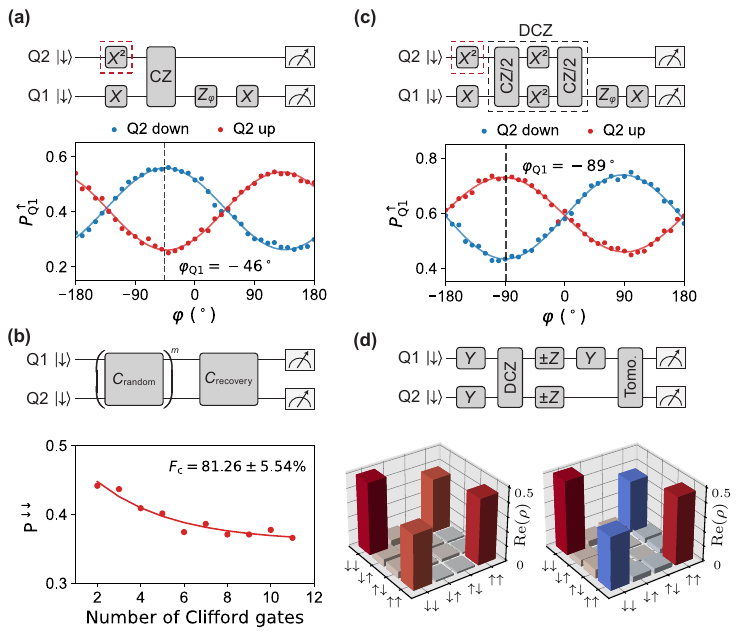}
  \caption{\label{fig4}Two-qubit logic gate and preparation of bell states. 
  (a) Spin-up probability of the target qubit (Q1) as a function of the phase of second $\pi/2$ pulse when the control qubit (Q2) is in spin-down (blue) and spin-up (red) states, measured using a Ramsey-type sequence. The red dashed box indicates an $X^2$ gate for preparing the control qubit in the spin-up state. The vertical dotted line indicates the single-qubit phase of Q1 required for a CZ gate.  
  (b) Two-qubit joint probability as a function of the number of two-qubit Clifford gates with an exponential fit (red line), measured using two-qubit RB sequences.  
  (c) Similar to (a), but measured using a Hahn echo-type sequence, with the single-qubit phase of Q1 centered near the theoretical value of $-90^{\circ}$. 
  (d) Upper: gate sequences to prepare Bell states and reconstruct the two-qubit density matrices. Lower: reconstructed density matrices (real part shown) of the $\ket{\Phi^+}$ and $\ket{\Phi^-}$ states, with fidelities of 84\% and 85\%, respectively.}
\end{figure*}

Next, we implement a two-qubit CZ gate, which causes the target qubit to undergo a $Z(\pi)$ evolution only when the control qubit is in the $\ket{\downarrow}$ state. To observe the spin-dependent phase evolution, we use a Ramsey-like sequence to measure the spin-up probability of the target qubit (Q1) by adjusting the relative phase of the second $\pi/2$ pulse. \figref{fig4}(a) shows the measured results when the control qubit (Q2) is prepared in the $\ket{\downarrow}$ or $\ket{\uparrow}$ states, respectively. The exchange pulse duration (80 ns at $J \approx 6.25$ MHz) has been calibrated to ensure that the two oscillation curves of Q1 are $180^{\circ}$ out of phase (see \supplementary). A CZ gate can be achieved by choosing the single-qubit phase of Q1 where the spin-up probability reaches its extreme point [dashed line in \autoref{fig4}(a)]. By switching the roles of Q1 and Q2, we can determine the single-qubit phase required for Q2 (see \supplementary).

The CZ gate performance is characterized using two-qubit RB.\cite{RN2045} Similar to single-qubit RB, randomly generated two-qubit Clifford gates are performed with the qubits prepared in the $\ket{\downarrow \downarrow}$ state, and the two-qubit joint probability is measured, as shown in \autoref{fig4}(b). By fitting the decay, we determine the average Clifford gate fidelity to be $F_{\text{2Q}}^{\text{c}} = (1+3p_{\text{ref}})/4 = 82.16 \pm 5.54\%$. Considering that each two-qubit Clifford gate, on average, comprises 8.25 single-qubit gates and 1.5 CZ gates,\cite{RN11996} we infer that the CZ gate fidelity is approximately 91\%, after eliminating the single-qubit gate errors. This result is comparable to the value extracted from character RB experiments.\cite{RN1413} Significant instability in the single-qubit phase is observed during the implementation of the CZ gate, primarily attributed to low-frequency noise from nuclear spin fluctuations and/or charge noise, which likely limits the CZ gate fidelity in our device. This instability also impedes benchmarking the CZ gate fidelity using two-qubit IRB.

To suppress single-qubit phase noise, dynamical decoupling pulses are introduced, as evidenced by the extended phase coherence times achieved using the Hahn echo sequence. Consequently, we decompose the CZ operation into two CZ/2 operations separated by an $X^2$ gate on each qubit, thereby implementing a decoupled CZ (DCZ) gate.\cite{RN1419, RN9013, RN11423, RN8962, RN11494} \figref{fig4}(c) shows the spin-dependent phase evolution for Q1, similar to the CZ gate, but with the spin-up probability extrema located near $\pm 90^\circ$. This indicates that the local single-qubit phases accumulated during the exchange pulses are effectively eliminated by the refocusing pulses, thereby mitigating the instability issues associated with local single-qubit phases.

To demonstrate the functionality of the two-qubit quantum device and provide a preliminary estimate of the fidelity of the DCZ gate, we present the preparation of Bell states. The upper panel in \autoref{fig4}(d) illustrates the gate sequences used for preparing the Bell states, including single-qubit gates and native two-qubit DCZ gates. Quantum state tomography is performed at the end of the sequences to evaluate the entanglement of the Bell states. By comparing the reconstructed density matrix with the ideal state using $F = \braket{\psi | \rho | \psi}$, we obtain an average Bell state fidelity of 85\% (91\%) and a concurrence of 0.73 (0.85) before (after) removing SPAM errors. SPAM errors are removed by following the procedure adopted from Ref. \onlinecite{RN1205} (see \supplementary). These results exceed previously reported values with a detuning pulse and are comparable to those with symmetry operation,\cite{RN9013} both in natural silicon. We attribute the improved fidelity primarily to the suppression of charge noise through symmetry operations. For a more comprehensive assessment, future studies could implement two-qubit RB protocols to assess the fidelity of DCZ gates.\cite{RN11494}

In conclusion, by utilizing advanced device structures, qubit manipulation techniques, and optimization methods, we have achieved single-qubit gate fidelities exceeding 99\%, a two-qubit CZ gate with a fidelity of 91\%, and the preparation of Bell states with a fidelity of 91\% using a natural Si/SiGe quantum dot device. These results initially demonstrate the potential for achieving high-fidelity gate operation in natural silicon, although there are limitations in our device that require further improvement. Firstly, the dephasing times in our device are shorter than those achieved with optimized micromagnets in natural silicon,\cite{RN1564} indicating that charge noise is a dominant source of low-frequency noise. This issue can be mitigated by optimizing the micromagnet structure.\cite{RN12049, RN10499} The single-qubit fidelities do not appear to be limited by the dephasing time given the longer $T_2^\text{Rabi}$, but are likely limited by crosstalk induced by residual coupling\cite{RN10511, RN9013} and systematic errors in control pulses\cite{RN1540}. For two-qubit gates, a larger exchange coupling is required to suppress the dephasing effects, which can be achievable by further employing virtual gates while operating at the charge-symmetry position. Additionally, nuclear spin noise, another major noise source, can be effectively suppressed using error mitigation schemes such as real-time Hamiltonian parameter estimation, as applied in GaAs.\cite{RN6097, RN6006, RN12039, RN11685} To achieve this, methods need to be developed to improve measurement bandwidth, such as radiofrequency reflectometry (RF).\cite{RN8623, RN8062} With these improvements, particularly in the suppression of charge and nuclear noise, we anticipate that it will be possible to achieve high-fidelity two-qubit gates even in non-purified silicon, which could advance semiconductor quantum computing using industry-standard natural silicon.

This work was supported by the Innovation Program for Quantum Science and Technology (Grant No. 2021ZD0302300) and the National Natural Science Foundation of China (Grants No. 92265113, No. 12074368, No. 12034018, and No. 92165207). Part of this work was carried out at the University of Science and Technology of China Center for Micro- and Nanoscale Research and Fabrication.

\bibliography{ref}

\begin{thebibliography}{45}%
\makeatletter
\providecommand \@ifxundefined [1]{%
 \@ifx{#1\undefined}
}%
\providecommand \@ifnum [1]{%
 \ifnum #1\expandafter \@firstoftwo
 \else \expandafter \@secondoftwo
 \fi
}%
\providecommand \@ifx [1]{%
 \ifx #1\expandafter \@firstoftwo
 \else \expandafter \@secondoftwo
 \fi
}%
\providecommand \natexlab [1]{#1}%
\providecommand \enquote  [1]{``#1''}%
\providecommand \bibnamefont  [1]{#1}%
\providecommand \bibfnamefont [1]{#1}%
\providecommand \citenamefont [1]{#1}%
\providecommand \href@noop [0]{\@secondoftwo}%
\providecommand \href [0]{\begingroup \@sanitize@url \@href}%
\providecommand \@href[1]{\@@startlink{#1}\@@href}%
\providecommand \@@href[1]{\endgroup#1\@@endlink}%
\providecommand \@sanitize@url [0]{\catcode `\\12\catcode `\$12\catcode `\&12\catcode `\#12\catcode `\^12\catcode `\_12\catcode `\%12\relax}%
\providecommand \@@startlink[1]{}%
\providecommand \@@endlink[0]{}%
\providecommand \url  [0]{\begingroup\@sanitize@url \@url }%
\providecommand \@url [1]{\endgroup\@href {#1}{\urlprefix }}%
\providecommand \urlprefix  [0]{URL }%
\providecommand \Eprint [0]{\href }%
\providecommand \doibase [0]{https://doi.org/}%
\providecommand \selectlanguage [0]{\@gobble}%
\providecommand \bibinfo  [0]{\@secondoftwo}%
\providecommand \bibfield  [0]{\@secondoftwo}%
\providecommand \translation [1]{[#1]}%
\providecommand \BibitemOpen [0]{}%
\providecommand \bibitemStop [0]{}%
\providecommand \bibitemNoStop [0]{.\EOS\space}%
\providecommand \EOS [0]{\spacefactor3000\relax}%
\providecommand \BibitemShut  [1]{\csname bibitem#1\endcsname}%
\let\auto@bib@innerbib\@empty
\bibitem [{\citenamefont {Wang}, \citenamefont {Fowler},\ and\ \citenamefont {Hollenberg}(2011)}]{RN10693}%
  \BibitemOpen
  \bibfield  {author} {\bibinfo {author} {\bibfnamefont {D.~S.}\ \bibnamefont {Wang}}, \bibinfo {author} {\bibfnamefont {A.~G.}\ \bibnamefont {Fowler}},\ and\ \bibinfo {author} {\bibfnamefont {L.~C.~L.}\ \bibnamefont {Hollenberg}},\ }\bibfield  {title} {\enquote {\bibinfo {title} {Surface code quantum computing with error rates over $1\%$},}\ }\href {https://doi.org/10.1103/PhysRevA.83.020302} {\bibfield  {journal} {\bibinfo  {journal} {Phys. Rev. A}\ }\textbf {\bibinfo {volume} {83}},\ \bibinfo {pages} {020302} (\bibinfo {year} {2011})}\BibitemShut {NoStop}%
\bibitem [{\citenamefont {Fowler}\ \emph {et~al.}(2012)\citenamefont {Fowler}, \citenamefont {Mariantoni}, \citenamefont {Martinis},\ and\ \citenamefont {Cleland}}]{RN11492}%
  \BibitemOpen
  \bibfield  {author} {\bibinfo {author} {\bibfnamefont {A.~G.}\ \bibnamefont {Fowler}}, \bibinfo {author} {\bibfnamefont {M.}~\bibnamefont {Mariantoni}}, \bibinfo {author} {\bibfnamefont {J.~M.}\ \bibnamefont {Martinis}},\ and\ \bibinfo {author} {\bibfnamefont {A.~N.}\ \bibnamefont {Cleland}},\ }\bibfield  {title} {\enquote {\bibinfo {title} {Surface codes: Towards practical large-scale quantum computation},}\ }\href {https://doi.org/10.1103/PhysRevA.86.032324} {\bibfield  {journal} {\bibinfo  {journal} {Phys. Rev. A}\ }\textbf {\bibinfo {volume} {86}},\ \bibinfo {pages} {032324} (\bibinfo {year} {2012})}\BibitemShut {NoStop}%
\bibitem [{\citenamefont {Veldhorst}\ \emph {et~al.}(2014)\citenamefont {Veldhorst}, \citenamefont {Hwang}, \citenamefont {Yang}, \citenamefont {Leenstra}, \citenamefont {de~Ronde}, \citenamefont {Dehollain}, \citenamefont {Muhonen}, \citenamefont {Hudson}, \citenamefont {Itoh}, \citenamefont {Morello},\ and\ \citenamefont {Dzurak}}]{RN1302}%
  \BibitemOpen
  \bibfield  {author} {\bibinfo {author} {\bibfnamefont {M.}~\bibnamefont {Veldhorst}}, \bibinfo {author} {\bibfnamefont {J.~C.}\ \bibnamefont {Hwang}}, \bibinfo {author} {\bibfnamefont {C.~H.}\ \bibnamefont {Yang}}, \bibinfo {author} {\bibfnamefont {A.~W.}\ \bibnamefont {Leenstra}}, \bibinfo {author} {\bibfnamefont {B.}~\bibnamefont {de~Ronde}}, \bibinfo {author} {\bibfnamefont {J.~P.}\ \bibnamefont {Dehollain}}, \bibinfo {author} {\bibfnamefont {J.~T.}\ \bibnamefont {Muhonen}}, \bibinfo {author} {\bibfnamefont {F.~E.}\ \bibnamefont {Hudson}}, \bibinfo {author} {\bibfnamefont {K.~M.}\ \bibnamefont {Itoh}}, \bibinfo {author} {\bibfnamefont {A.}~\bibnamefont {Morello}},\ and\ \bibinfo {author} {\bibfnamefont {A.~S.}\ \bibnamefont {Dzurak}},\ }\bibfield  {title} {\enquote {\bibinfo {title} {An addressable quantum dot qubit with fault-tolerant control-fidelity},}\ }\href {https://doi.org/10.1038/nnano.2014.216} {\bibfield  {journal} {\bibinfo  {journal} {Nat. Nanotechnol.}\ }\textbf {\bibinfo {volume} {9}},\ \bibinfo {pages} {981--985} (\bibinfo {year} {2014})}\BibitemShut {NoStop}%
\bibitem [{\citenamefont {Zhang}\ \emph {et~al.}(2019)\citenamefont {Zhang}, \citenamefont {Li}, \citenamefont {Cao}, \citenamefont {Xiao}, \citenamefont {Guo},\ and\ \citenamefont {Guo}}]{RN7360}%
  \BibitemOpen
  \bibfield  {author} {\bibinfo {author} {\bibfnamefont {X.}~\bibnamefont {Zhang}}, \bibinfo {author} {\bibfnamefont {H.~O.}\ \bibnamefont {Li}}, \bibinfo {author} {\bibfnamefont {G.}~\bibnamefont {Cao}}, \bibinfo {author} {\bibfnamefont {M.}~\bibnamefont {Xiao}}, \bibinfo {author} {\bibfnamefont {G.~C.}\ \bibnamefont {Guo}},\ and\ \bibinfo {author} {\bibfnamefont {G.~P.}\ \bibnamefont {Guo}},\ }\bibfield  {title} {\enquote {\bibinfo {title} {Semiconductor quantum computation},}\ }\href {https://doi.org/10.1093/nsr/nwy153} {\bibfield  {journal} {\bibinfo  {journal} {Natl. Sci. Rev.}\ }\textbf {\bibinfo {volume} {6}},\ \bibinfo {pages} {32--54} (\bibinfo {year} {2019})}\BibitemShut {NoStop}%
\bibitem [{\citenamefont {Liu}\ \emph {et~al.}(2022)\citenamefont {Liu}, \citenamefont {Wang}, \citenamefont {Wang}, \citenamefont {Sun}, \citenamefont {Yin}, \citenamefont {Li}, \citenamefont {Cao},\ and\ \citenamefont {Guo}}]{RN11424}%
  \BibitemOpen
  \bibfield  {author} {\bibinfo {author} {\bibfnamefont {H.-W.}\ \bibnamefont {Liu}}, \bibinfo {author} {\bibfnamefont {B.-C.}\ \bibnamefont {Wang}}, \bibinfo {author} {\bibfnamefont {N.}~\bibnamefont {Wang}}, \bibinfo {author} {\bibfnamefont {Z.-H.}\ \bibnamefont {Sun}}, \bibinfo {author} {\bibfnamefont {H.-L.}\ \bibnamefont {Yin}}, \bibinfo {author} {\bibfnamefont {H.-O.}\ \bibnamefont {Li}}, \bibinfo {author} {\bibfnamefont {G.}~\bibnamefont {Cao}},\ and\ \bibinfo {author} {\bibfnamefont {G.-P.}\ \bibnamefont {Guo}},\ }\bibfield  {title} {\enquote {\bibinfo {title} {An automated approach for consecutive tuning of quantum dot arrays},}\ }\href {https://doi.org/10.1063/5.0111128} {\bibfield  {journal} {\bibinfo  {journal} {Appl. Phys. Lett.}\ }\textbf {\bibinfo {volume} {121}},\ \bibinfo {pages} {084002} (\bibinfo {year} {2022})}\BibitemShut {NoStop}%
\bibitem [{\citenamefont {Gonzalez-Zalba}\ \emph {et~al.}(2021)\citenamefont {Gonzalez-Zalba}, \citenamefont {de~Franceschi}, \citenamefont {Charbon}, \citenamefont {Meunier}, \citenamefont {Vinet},\ and\ \citenamefont {Dzurak}}]{RN11647}%
  \BibitemOpen
  \bibfield  {author} {\bibinfo {author} {\bibfnamefont {M.~F.}\ \bibnamefont {Gonzalez-Zalba}}, \bibinfo {author} {\bibfnamefont {S.}~\bibnamefont {de~Franceschi}}, \bibinfo {author} {\bibfnamefont {E.}~\bibnamefont {Charbon}}, \bibinfo {author} {\bibfnamefont {T.}~\bibnamefont {Meunier}}, \bibinfo {author} {\bibfnamefont {M.}~\bibnamefont {Vinet}},\ and\ \bibinfo {author} {\bibfnamefont {A.~S.}\ \bibnamefont {Dzurak}},\ }\bibfield  {title} {\enquote {\bibinfo {title} {Scaling silicon-based quantum computing using $\text{CMOS}$ technology},}\ }\href {https://doi.org/10.1038/s41928-021-00681-y} {\bibfield  {journal} {\bibinfo  {journal} {Nat. Electron.}\ }\textbf {\bibinfo {volume} {4}},\ \bibinfo {pages} {872--884} (\bibinfo {year} {2021})}\BibitemShut {NoStop}%
\bibitem [{\citenamefont {Zwerver}\ \emph {et~al.}(2022)\citenamefont {Zwerver}, \citenamefont {Krähenmann}, \citenamefont {Watson}, \citenamefont {Lampert}, \citenamefont {George}, \citenamefont {Pillarisetty}, \citenamefont {Bojarski}, \citenamefont {Amin}, \citenamefont {Amitonov}, \citenamefont {Boter}, \citenamefont {Caudillo}, \citenamefont {Correas-Serrano}, \citenamefont {Dehollain}, \citenamefont {Droulers}, \citenamefont {Henry}, \citenamefont {Kotlyar}, \citenamefont {Lodari}, \citenamefont {Lüthi}, \citenamefont {Michalak}, \citenamefont {Mueller}, \citenamefont {Neyens}, \citenamefont {Roberts}, \citenamefont {Samkharadze}, \citenamefont {Zheng}, \citenamefont {Zietz}, \citenamefont {Scappucci}, \citenamefont {Veldhorst}, \citenamefont {Vandersypen},\ and\ \citenamefont {Clarke}}]{RN10696}%
  \BibitemOpen
  \bibfield  {author} {\bibinfo {author} {\bibfnamefont {A.~M.~J.}\ \bibnamefont {Zwerver}}, \bibinfo {author} {\bibfnamefont {T.}~\bibnamefont {Krähenmann}}, \bibinfo {author} {\bibfnamefont {T.~F.}\ \bibnamefont {Watson}}, \bibinfo {author} {\bibfnamefont {L.}~\bibnamefont {Lampert}}, \bibinfo {author} {\bibfnamefont {H.~C.}\ \bibnamefont {George}}, \bibinfo {author} {\bibfnamefont {R.}~\bibnamefont {Pillarisetty}}, \bibinfo {author} {\bibfnamefont {S.~A.}\ \bibnamefont {Bojarski}}, \bibinfo {author} {\bibfnamefont {P.}~\bibnamefont {Amin}}, \bibinfo {author} {\bibfnamefont {S.~V.}\ \bibnamefont {Amitonov}}, \bibinfo {author} {\bibfnamefont {J.~M.}\ \bibnamefont {Boter}}, \bibinfo {author} {\bibfnamefont {R.}~\bibnamefont {Caudillo}}, \bibinfo {author} {\bibfnamefont {D.}~\bibnamefont {Correas-Serrano}}, \bibinfo {author} {\bibfnamefont {J.~P.}\ \bibnamefont {Dehollain}}, \bibinfo {author} {\bibfnamefont {G.}~\bibnamefont {Droulers}}, \bibinfo {author} {\bibfnamefont {E.~M.}\ \bibnamefont {Henry}}, \bibinfo {author} {\bibfnamefont {R.}~\bibnamefont {Kotlyar}}, \bibinfo {author} {\bibfnamefont {M.}~\bibnamefont {Lodari}}, \bibinfo {author} {\bibfnamefont {F.}~\bibnamefont {Lüthi}}, \bibinfo {author} {\bibfnamefont {D.~J.}\ \bibnamefont {Michalak}}, \bibinfo {author} {\bibfnamefont {B.~K.}\ \bibnamefont {Mueller}}, \bibinfo {author} {\bibfnamefont {S.}~\bibnamefont {Neyens}}, \bibinfo {author} {\bibfnamefont {J.}~\bibnamefont {Roberts}}, \bibinfo {author} {\bibfnamefont {N.}~\bibnamefont {Samkharadze}}, \bibinfo {author} {\bibfnamefont {G.}~\bibnamefont {Zheng}}, \bibinfo {author} {\bibfnamefont {O.~K.}\ \bibnamefont {Zietz}}, \bibinfo {author} {\bibfnamefont {G.}~\bibnamefont {Scappucci}}, \bibinfo {author} {\bibfnamefont {M.}~\bibnamefont {Veldhorst}}, \bibinfo {author} {\bibfnamefont {L.~M.~K.}\ \bibnamefont {Vandersypen}},\ and\ \bibinfo {author} {\bibfnamefont {J.~S.}\ \bibnamefont {Clarke}},\ }\bibfield  {title} {\enquote {\bibinfo {title} {Qubits made by advanced semiconductor manufacturing},}\ }\href {https://doi.org/10.1038/s41928-022-00727-9} {\bibfield  {journal} {\bibinfo  {journal} {Nat. Electron.}\ }\textbf {\bibinfo {volume} {5}},\ \bibinfo {pages} {184--190} (\bibinfo {year} {2022})}\BibitemShut {NoStop}%
\bibitem [{\citenamefont {Neyens}\ \emph {et~al.}(2024)\citenamefont {Neyens}, \citenamefont {Zietz}, \citenamefont {Watson}, \citenamefont {Luthi}, \citenamefont {Nethwewala}, \citenamefont {George}, \citenamefont {Henry}, \citenamefont {Islam}, \citenamefont {Wagner}, \citenamefont {Borjans}, \citenamefont {Connors}, \citenamefont {Corrigan}, \citenamefont {Curry}, \citenamefont {Keith}, \citenamefont {Kotlyar}, \citenamefont {Lampert}, \citenamefont {Madzik}, \citenamefont {Millard}, \citenamefont {Mohiyaddin}, \citenamefont {Pellerano}, \citenamefont {Pillarisetty}, \citenamefont {Ramsey}, \citenamefont {Savytskyy}, \citenamefont {Schaal}, \citenamefont {Zheng}, \citenamefont {Ziegler}, \citenamefont {Bishop}, \citenamefont {Bojarski}, \citenamefont {Roberts},\ and\ \citenamefont {Clarke}}]{RN11981}%
  \BibitemOpen
  \bibfield  {author} {\bibinfo {author} {\bibfnamefont {S.}~\bibnamefont {Neyens}}, \bibinfo {author} {\bibfnamefont {O.~K.}\ \bibnamefont {Zietz}}, \bibinfo {author} {\bibfnamefont {T.~F.}\ \bibnamefont {Watson}}, \bibinfo {author} {\bibfnamefont {F.}~\bibnamefont {Luthi}}, \bibinfo {author} {\bibfnamefont {A.}~\bibnamefont {Nethwewala}}, \bibinfo {author} {\bibfnamefont {H.~C.}\ \bibnamefont {George}}, \bibinfo {author} {\bibfnamefont {E.}~\bibnamefont {Henry}}, \bibinfo {author} {\bibfnamefont {M.}~\bibnamefont {Islam}}, \bibinfo {author} {\bibfnamefont {A.~J.}\ \bibnamefont {Wagner}}, \bibinfo {author} {\bibfnamefont {F.}~\bibnamefont {Borjans}}, \bibinfo {author} {\bibfnamefont {E.~J.}\ \bibnamefont {Connors}}, \bibinfo {author} {\bibfnamefont {J.}~\bibnamefont {Corrigan}}, \bibinfo {author} {\bibfnamefont {M.~J.}\ \bibnamefont {Curry}}, \bibinfo {author} {\bibfnamefont {D.}~\bibnamefont {Keith}}, \bibinfo {author} {\bibfnamefont {R.}~\bibnamefont {Kotlyar}}, \bibinfo {author} {\bibfnamefont {L.~F.}\ \bibnamefont {Lampert}}, \bibinfo {author} {\bibfnamefont {M.~T.}\ \bibnamefont {Madzik}}, \bibinfo {author} {\bibfnamefont {K.}~\bibnamefont {Millard}}, \bibinfo {author} {\bibfnamefont {F.~A.}\ \bibnamefont {Mohiyaddin}}, \bibinfo {author} {\bibfnamefont {S.}~\bibnamefont {Pellerano}}, \bibinfo {author} {\bibfnamefont {R.}~\bibnamefont {Pillarisetty}}, \bibinfo {author} {\bibfnamefont {M.}~\bibnamefont {Ramsey}}, \bibinfo {author} {\bibfnamefont {R.}~\bibnamefont {Savytskyy}}, \bibinfo {author} {\bibfnamefont {S.}~\bibnamefont {Schaal}}, \bibinfo {author} {\bibfnamefont {G.}~\bibnamefont {Zheng}}, \bibinfo {author} {\bibfnamefont {J.}~\bibnamefont {Ziegler}}, \bibinfo {author} {\bibfnamefont {N.~C.}\ \bibnamefont {Bishop}}, \bibinfo {author} {\bibfnamefont {S.}~\bibnamefont {Bojarski}}, \bibinfo {author} {\bibfnamefont {J.}~\bibnamefont {Roberts}},\ and\ \bibinfo {author} {\bibfnamefont {J.~S.}\ \bibnamefont {Clarke}},\ }\bibfield  {title} {\enquote {\bibinfo {title} {Probing single electrons across 300-mm spin qubit wafers},}\ }\href {https://doi.org/10.1038/s41586-024-07275-6} {\bibfield  {journal} {\bibinfo  {journal} {Nature}\ }\textbf {\bibinfo {volume} {629}},\ \bibinfo {pages} {80--85} (\bibinfo {year} {2024})}\BibitemShut {NoStop}%
\bibitem [{\citenamefont {Vandersypen}\ and\ \citenamefont {Eriksson}(2019)}]{RN4285}%
  \BibitemOpen
  \bibfield  {author} {\bibinfo {author} {\bibfnamefont {L.~M.~K.}\ \bibnamefont {Vandersypen}}\ and\ \bibinfo {author} {\bibfnamefont {M.~A.}\ \bibnamefont {Eriksson}},\ }\bibfield  {title} {\enquote {\bibinfo {title} {Quantum computing with semiconductor spins},}\ }\href {https://doi.org/10.1063/Pt.3.4270} {\bibfield  {journal} {\bibinfo  {journal} {Phys. Today}\ }\textbf {\bibinfo {volume} {72}},\ \bibinfo {pages} {38--45} (\bibinfo {year} {2019})}\BibitemShut {NoStop}%
\bibitem [{\citenamefont {Kawakami}\ \emph {et~al.}(2014)\citenamefont {Kawakami}, \citenamefont {Scarlino}, \citenamefont {Ward}, \citenamefont {Braakman}, \citenamefont {Savage}, \citenamefont {Lagally}, \citenamefont {Friesen}, \citenamefont {Coppersmith}, \citenamefont {Eriksson},\ and\ \citenamefont {Vandersypen}}]{RN4323}%
  \BibitemOpen
  \bibfield  {author} {\bibinfo {author} {\bibfnamefont {E.}~\bibnamefont {Kawakami}}, \bibinfo {author} {\bibfnamefont {P.}~\bibnamefont {Scarlino}}, \bibinfo {author} {\bibfnamefont {D.~R.}\ \bibnamefont {Ward}}, \bibinfo {author} {\bibfnamefont {F.~R.}\ \bibnamefont {Braakman}}, \bibinfo {author} {\bibfnamefont {D.~E.}\ \bibnamefont {Savage}}, \bibinfo {author} {\bibfnamefont {M.~G.}\ \bibnamefont {Lagally}}, \bibinfo {author} {\bibfnamefont {M.}~\bibnamefont {Friesen}}, \bibinfo {author} {\bibfnamefont {S.~N.}\ \bibnamefont {Coppersmith}}, \bibinfo {author} {\bibfnamefont {M.~A.}\ \bibnamefont {Eriksson}},\ and\ \bibinfo {author} {\bibfnamefont {L.~M.~K.}\ \bibnamefont {Vandersypen}},\ }\bibfield  {title} {\enquote {\bibinfo {title} {Electrical control of a long-lived spin qubit in a $\text{Si/SiGe}$ quantum dot},}\ }\href {https://doi.org/10.1038/Nnano.2014.153} {\bibfield  {journal} {\bibinfo  {journal} {Nat. Nanotechnol.}\ }\textbf {\bibinfo {volume} {9}},\ \bibinfo {pages} {666--670} (\bibinfo {year} {2014})}\BibitemShut {NoStop}%
\bibitem [{\citenamefont {Kawakami}\ \emph {et~al.}(2016)\citenamefont {Kawakami}, \citenamefont {Jullien}, \citenamefont {Scarlino}, \citenamefont {Ward}, \citenamefont {Savage}, \citenamefont {Lagally}, \citenamefont {Dobrovitski}, \citenamefont {Friesen}, \citenamefont {Coppersmith}, \citenamefont {Eriksson},\ and\ \citenamefont {Vandersypen}}]{RN1441}%
  \BibitemOpen
  \bibfield  {author} {\bibinfo {author} {\bibfnamefont {E.}~\bibnamefont {Kawakami}}, \bibinfo {author} {\bibfnamefont {T.}~\bibnamefont {Jullien}}, \bibinfo {author} {\bibfnamefont {P.}~\bibnamefont {Scarlino}}, \bibinfo {author} {\bibfnamefont {D.~R.}\ \bibnamefont {Ward}}, \bibinfo {author} {\bibfnamefont {D.~E.}\ \bibnamefont {Savage}}, \bibinfo {author} {\bibfnamefont {M.~G.}\ \bibnamefont {Lagally}}, \bibinfo {author} {\bibfnamefont {V.~V.}\ \bibnamefont {Dobrovitski}}, \bibinfo {author} {\bibfnamefont {M.}~\bibnamefont {Friesen}}, \bibinfo {author} {\bibfnamefont {S.~N.}\ \bibnamefont {Coppersmith}}, \bibinfo {author} {\bibfnamefont {M.~A.}\ \bibnamefont {Eriksson}},\ and\ \bibinfo {author} {\bibfnamefont {L.~M.~K.}\ \bibnamefont {Vandersypen}},\ }\bibfield  {title} {\enquote {\bibinfo {title} {Gate fidelity and coherence of an electron spin in an $\text{Si/SiGe}$ quantum dot with micromagnet},}\ }\href {https://doi.org/10.1073/pnas.1603251113} {\bibfield  {journal} {\bibinfo  {journal} {PNAS}\ }\textbf {\bibinfo {volume} {113}},\ \bibinfo {pages} {11738--11743} (\bibinfo {year} {2016})}\BibitemShut {NoStop}%
\bibitem [{\citenamefont {Takeda}\ \emph {et~al.}(2016)\citenamefont {Takeda}, \citenamefont {Kamioka}, \citenamefont {Otsuka}, \citenamefont {Yoneda}, \citenamefont {Nakajima}, \citenamefont {Delbecq}, \citenamefont {Amaha}, \citenamefont {Allison}, \citenamefont {Kodera}, \citenamefont {Oda},\ and\ \citenamefont {Tarucha}}]{RN1564}%
  \BibitemOpen
  \bibfield  {author} {\bibinfo {author} {\bibfnamefont {K.}~\bibnamefont {Takeda}}, \bibinfo {author} {\bibfnamefont {J.}~\bibnamefont {Kamioka}}, \bibinfo {author} {\bibfnamefont {T.}~\bibnamefont {Otsuka}}, \bibinfo {author} {\bibfnamefont {J.}~\bibnamefont {Yoneda}}, \bibinfo {author} {\bibfnamefont {T.}~\bibnamefont {Nakajima}}, \bibinfo {author} {\bibfnamefont {M.~R.}\ \bibnamefont {Delbecq}}, \bibinfo {author} {\bibfnamefont {S.}~\bibnamefont {Amaha}}, \bibinfo {author} {\bibfnamefont {G.}~\bibnamefont {Allison}}, \bibinfo {author} {\bibfnamefont {T.}~\bibnamefont {Kodera}}, \bibinfo {author} {\bibfnamefont {S.}~\bibnamefont {Oda}},\ and\ \bibinfo {author} {\bibfnamefont {S.}~\bibnamefont {Tarucha}},\ }\bibfield  {title} {\enquote {\bibinfo {title} {A fault-tolerant addressable spin qubit in a natural silicon quantum dot},}\ }\href {https://doi.org/10.1126/sciadv.1600694} {\bibfield  {journal} {\bibinfo  {journal} {Sci. Adv.}\ }\textbf {\bibinfo {volume} {2}},\ \bibinfo {pages} {e1600694} (\bibinfo {year} {2016})}\BibitemShut {NoStop}%
\bibitem [{\citenamefont {Watson}\ \emph {et~al.}(2018)\citenamefont {Watson}, \citenamefont {Philips}, \citenamefont {Kawakami}, \citenamefont {Ward}, \citenamefont {Scarlino}, \citenamefont {Veldhorst}, \citenamefont {Savage}, \citenamefont {Lagally}, \citenamefont {Friesen}, \citenamefont {Coppersmith}, \citenamefont {Eriksson},\ and\ \citenamefont {Vandersypen}}]{RN1419}%
  \BibitemOpen
  \bibfield  {author} {\bibinfo {author} {\bibfnamefont {T.~F.}\ \bibnamefont {Watson}}, \bibinfo {author} {\bibfnamefont {S.~G.~J.}\ \bibnamefont {Philips}}, \bibinfo {author} {\bibfnamefont {E.}~\bibnamefont {Kawakami}}, \bibinfo {author} {\bibfnamefont {D.~R.}\ \bibnamefont {Ward}}, \bibinfo {author} {\bibfnamefont {P.}~\bibnamefont {Scarlino}}, \bibinfo {author} {\bibfnamefont {M.}~\bibnamefont {Veldhorst}}, \bibinfo {author} {\bibfnamefont {D.~E.}\ \bibnamefont {Savage}}, \bibinfo {author} {\bibfnamefont {M.~G.}\ \bibnamefont {Lagally}}, \bibinfo {author} {\bibfnamefont {M.}~\bibnamefont {Friesen}}, \bibinfo {author} {\bibfnamefont {S.~N.}\ \bibnamefont {Coppersmith}}, \bibinfo {author} {\bibfnamefont {M.~A.}\ \bibnamefont {Eriksson}},\ and\ \bibinfo {author} {\bibfnamefont {L.~M.~K.}\ \bibnamefont {Vandersypen}},\ }\bibfield  {title} {\enquote {\bibinfo {title} {A programmable two-qubit quantum processor in silicon},}\ }\href {https://doi.org/10.1038/nature25766} {\bibfield  {journal} {\bibinfo  {journal} {Nature}\ }\textbf {\bibinfo {volume} {555}},\ \bibinfo {pages} {633–637} (\bibinfo {year} {2018})}\BibitemShut {NoStop}%
\bibitem [{\citenamefont {Xue}\ \emph {et~al.}(2019)\citenamefont {Xue}, \citenamefont {Watson}, \citenamefont {Helsen}, \citenamefont {Ward}, \citenamefont {Savage}, \citenamefont {Lagally}, \citenamefont {Coppersmith}, \citenamefont {Eriksson}, \citenamefont {Wehner},\ and\ \citenamefont {Vandersypen}}]{RN1413}%
  \BibitemOpen
  \bibfield  {author} {\bibinfo {author} {\bibfnamefont {X.}~\bibnamefont {Xue}}, \bibinfo {author} {\bibfnamefont {T.~F.}\ \bibnamefont {Watson}}, \bibinfo {author} {\bibfnamefont {J.}~\bibnamefont {Helsen}}, \bibinfo {author} {\bibfnamefont {D.~R.}\ \bibnamefont {Ward}}, \bibinfo {author} {\bibfnamefont {D.~E.}\ \bibnamefont {Savage}}, \bibinfo {author} {\bibfnamefont {M.~G.}\ \bibnamefont {Lagally}}, \bibinfo {author} {\bibfnamefont {S.~N.}\ \bibnamefont {Coppersmith}}, \bibinfo {author} {\bibfnamefont {M.~A.}\ \bibnamefont {Eriksson}}, \bibinfo {author} {\bibfnamefont {S.}~\bibnamefont {Wehner}},\ and\ \bibinfo {author} {\bibfnamefont {L.~M.~K.}\ \bibnamefont {Vandersypen}},\ }\bibfield  {title} {\enquote {\bibinfo {title} {Benchmarking gate fidelities in a $\text{Si/SiGe}$ two-qubit device},}\ }\href {https://doi.org/10.1103/PhysRevX.9.021011} {\bibfield  {journal} {\bibinfo  {journal} {Phys. Rev. X}\ }\textbf {\bibinfo {volume} {9}},\ \bibinfo {pages} {021011} (\bibinfo {year} {2019})}\BibitemShut {NoStop}%
\bibitem [{\citenamefont {Xue}\ \emph {et~al.}(2021)\citenamefont {Xue}, \citenamefont {Patra}, \citenamefont {van Dijk}, \citenamefont {Samkharadze}, \citenamefont {Subramanian}, \citenamefont {Corna}, \citenamefont {Paquelet~Wuetz}, \citenamefont {Jeon}, \citenamefont {Sheikh}, \citenamefont {Juarez-Hernandez}, \citenamefont {Esparza}, \citenamefont {Rampurawala}, \citenamefont {Carlton}, \citenamefont {Ravikumar}, \citenamefont {Nieva}, \citenamefont {Kim}, \citenamefont {Lee}, \citenamefont {Sammak}, \citenamefont {Scappucci}, \citenamefont {Veldhorst}, \citenamefont {Sebastiano}, \citenamefont {Babaie}, \citenamefont {Pellerano}, \citenamefont {Charbon},\ and\ \citenamefont {Vandersypen}}]{RN9045}%
  \BibitemOpen
  \bibfield  {author} {\bibinfo {author} {\bibfnamefont {X.}~\bibnamefont {Xue}}, \bibinfo {author} {\bibfnamefont {B.}~\bibnamefont {Patra}}, \bibinfo {author} {\bibfnamefont {J.~P.~G.}\ \bibnamefont {van Dijk}}, \bibinfo {author} {\bibfnamefont {N.}~\bibnamefont {Samkharadze}}, \bibinfo {author} {\bibfnamefont {S.}~\bibnamefont {Subramanian}}, \bibinfo {author} {\bibfnamefont {A.}~\bibnamefont {Corna}}, \bibinfo {author} {\bibfnamefont {B.}~\bibnamefont {Paquelet~Wuetz}}, \bibinfo {author} {\bibfnamefont {C.}~\bibnamefont {Jeon}}, \bibinfo {author} {\bibfnamefont {F.}~\bibnamefont {Sheikh}}, \bibinfo {author} {\bibfnamefont {E.}~\bibnamefont {Juarez-Hernandez}}, \bibinfo {author} {\bibfnamefont {B.~P.}\ \bibnamefont {Esparza}}, \bibinfo {author} {\bibfnamefont {H.}~\bibnamefont {Rampurawala}}, \bibinfo {author} {\bibfnamefont {B.}~\bibnamefont {Carlton}}, \bibinfo {author} {\bibfnamefont {S.}~\bibnamefont {Ravikumar}}, \bibinfo {author} {\bibfnamefont {C.}~\bibnamefont {Nieva}}, \bibinfo {author} {\bibfnamefont {S.}~\bibnamefont {Kim}}, \bibinfo {author} {\bibfnamefont {H.~J.}\ \bibnamefont {Lee}}, \bibinfo {author} {\bibfnamefont {A.}~\bibnamefont {Sammak}}, \bibinfo {author} {\bibfnamefont {G.}~\bibnamefont {Scappucci}}, \bibinfo {author} {\bibfnamefont {M.}~\bibnamefont {Veldhorst}}, \bibinfo {author} {\bibfnamefont {F.}~\bibnamefont {Sebastiano}}, \bibinfo {author} {\bibfnamefont {M.}~\bibnamefont {Babaie}}, \bibinfo {author} {\bibfnamefont {S.}~\bibnamefont {Pellerano}}, \bibinfo {author} {\bibfnamefont {E.}~\bibnamefont {Charbon}},\ and\ \bibinfo {author} {\bibfnamefont {L.~M.~K.}\ \bibnamefont {Vandersypen}},\ }\bibfield  {title} {\enquote {\bibinfo {title} {$\text{CMOS}$-based cryogenic control of silicon quantum circuits},}\ }\href {https://doi.org/10.1038/s41586-021-03469-4} {\bibfield  {journal} {\bibinfo  {journal} {Nature}\ }\textbf {\bibinfo {volume} {593}},\ \bibinfo {pages} {205--210} (\bibinfo {year} {2021})}\BibitemShut {NoStop}%
\bibitem [{\citenamefont {Noiri}\ \emph {et~al.}(2022{\natexlab{a}})\citenamefont {Noiri}, \citenamefont {Takeda}, \citenamefont {Nakajima}, \citenamefont {Kobayashi}, \citenamefont {Sammak}, \citenamefont {Scappucci},\ and\ \citenamefont {Tarucha}}]{RN10510}%
  \BibitemOpen
  \bibfield  {author} {\bibinfo {author} {\bibfnamefont {A.}~\bibnamefont {Noiri}}, \bibinfo {author} {\bibfnamefont {K.}~\bibnamefont {Takeda}}, \bibinfo {author} {\bibfnamefont {T.}~\bibnamefont {Nakajima}}, \bibinfo {author} {\bibfnamefont {T.}~\bibnamefont {Kobayashi}}, \bibinfo {author} {\bibfnamefont {A.}~\bibnamefont {Sammak}}, \bibinfo {author} {\bibfnamefont {G.}~\bibnamefont {Scappucci}},\ and\ \bibinfo {author} {\bibfnamefont {S.}~\bibnamefont {Tarucha}},\ }\bibfield  {title} {\enquote {\bibinfo {title} {Fast universal quantum gate above the fault-tolerance threshold in silicon},}\ }\href {https://doi.org/10.1038/s41586-021-04182-y} {\bibfield  {journal} {\bibinfo  {journal} {Nature}\ }\textbf {\bibinfo {volume} {601}},\ \bibinfo {pages} {338--342} (\bibinfo {year} {2022}{\natexlab{a}})}\BibitemShut {NoStop}%
\bibitem [{\citenamefont {Xue}\ \emph {et~al.}(2022)\citenamefont {Xue}, \citenamefont {Russ}, \citenamefont {Samkharadze}, \citenamefont {Undseth}, \citenamefont {Sammak}, \citenamefont {Scappucci},\ and\ \citenamefont {Vandersypen}}]{RN10511}%
  \BibitemOpen
  \bibfield  {author} {\bibinfo {author} {\bibfnamefont {X.}~\bibnamefont {Xue}}, \bibinfo {author} {\bibfnamefont {M.}~\bibnamefont {Russ}}, \bibinfo {author} {\bibfnamefont {N.}~\bibnamefont {Samkharadze}}, \bibinfo {author} {\bibfnamefont {B.}~\bibnamefont {Undseth}}, \bibinfo {author} {\bibfnamefont {A.}~\bibnamefont {Sammak}}, \bibinfo {author} {\bibfnamefont {G.}~\bibnamefont {Scappucci}},\ and\ \bibinfo {author} {\bibfnamefont {L.~M.~K.}\ \bibnamefont {Vandersypen}},\ }\bibfield  {title} {\enquote {\bibinfo {title} {Quantum logic with spin qubits crossing the surface code threshold},}\ }\href {https://doi.org/10.1038/s41586-021-04273-w} {\bibfield  {journal} {\bibinfo  {journal} {Nature}\ }\textbf {\bibinfo {volume} {601}},\ \bibinfo {pages} {343--347} (\bibinfo {year} {2022})}\BibitemShut {NoStop}%
\bibitem [{\citenamefont {Mills}\ \emph {et~al.}(2022)\citenamefont {Mills}, \citenamefont {Guinn}, \citenamefont {Gullans}, \citenamefont {Sigillito}, \citenamefont {Feldman}, \citenamefont {Nielsen},\ and\ \citenamefont {Petta}}]{RN10653}%
  \BibitemOpen
  \bibfield  {author} {\bibinfo {author} {\bibfnamefont {A.~R.}\ \bibnamefont {Mills}}, \bibinfo {author} {\bibfnamefont {C.~R.}\ \bibnamefont {Guinn}}, \bibinfo {author} {\bibfnamefont {M.~J.}\ \bibnamefont {Gullans}}, \bibinfo {author} {\bibfnamefont {A.~J.}\ \bibnamefont {Sigillito}}, \bibinfo {author} {\bibfnamefont {M.~M.}\ \bibnamefont {Feldman}}, \bibinfo {author} {\bibfnamefont {E.}~\bibnamefont {Nielsen}},\ and\ \bibinfo {author} {\bibfnamefont {J.~R.}\ \bibnamefont {Petta}},\ }\bibfield  {title} {\enquote {\bibinfo {title} {Two-qubit silicon quantum processor with operation fidelity exceeding $99\%$},}\ }\href {https://doi.org/10.1126/sciadv.abn5130} {\bibfield  {journal} {\bibinfo  {journal} {Sci. Adv.}\ }\textbf {\bibinfo {volume} {8}},\ \bibinfo {pages} {eabn5130} (\bibinfo {year} {2022})}\BibitemShut {NoStop}%
\bibitem [{\citenamefont {Philips}\ \emph {et~al.}(2022)\citenamefont {Philips}, \citenamefont {Madzik}, \citenamefont {Amitonov}, \citenamefont {de~Snoo}, \citenamefont {Russ}, \citenamefont {Kalhor}, \citenamefont {Volk}, \citenamefont {Lawrie}, \citenamefont {Brousse}, \citenamefont {Tryputen}, \citenamefont {Wuetz}, \citenamefont {Sammak}, \citenamefont {Veldhorst}, \citenamefont {Scappucci},\ and\ \citenamefont {Vandersypen}}]{RN11493}%
  \BibitemOpen
  \bibfield  {author} {\bibinfo {author} {\bibfnamefont {S.~G.~J.}\ \bibnamefont {Philips}}, \bibinfo {author} {\bibfnamefont {M.~T.}\ \bibnamefont {Madzik}}, \bibinfo {author} {\bibfnamefont {S.~V.}\ \bibnamefont {Amitonov}}, \bibinfo {author} {\bibfnamefont {S.~L.}\ \bibnamefont {de~Snoo}}, \bibinfo {author} {\bibfnamefont {M.}~\bibnamefont {Russ}}, \bibinfo {author} {\bibfnamefont {N.}~\bibnamefont {Kalhor}}, \bibinfo {author} {\bibfnamefont {C.}~\bibnamefont {Volk}}, \bibinfo {author} {\bibfnamefont {W.~I.~L.}\ \bibnamefont {Lawrie}}, \bibinfo {author} {\bibfnamefont {D.}~\bibnamefont {Brousse}}, \bibinfo {author} {\bibfnamefont {L.}~\bibnamefont {Tryputen}}, \bibinfo {author} {\bibfnamefont {B.~P.}\ \bibnamefont {Wuetz}}, \bibinfo {author} {\bibfnamefont {A.}~\bibnamefont {Sammak}}, \bibinfo {author} {\bibfnamefont {M.}~\bibnamefont {Veldhorst}}, \bibinfo {author} {\bibfnamefont {G.}~\bibnamefont {Scappucci}},\ and\ \bibinfo {author} {\bibfnamefont {L.~M.~K.}\ \bibnamefont {Vandersypen}},\ }\bibfield  {title} {\enquote {\bibinfo {title} {Universal control of a six-qubit quantum processor in silicon},}\ }\href {https://doi.org/10.1038/s41586-022-05117-x} {\bibfield  {journal} {\bibinfo  {journal} {Nature}\ }\textbf {\bibinfo {volume} {609}},\ \bibinfo {pages} {919--924} (\bibinfo {year} {2022})}\BibitemShut {NoStop}%
\bibitem [{\citenamefont {Tokura}\ \emph {et~al.}(2006)\citenamefont {Tokura}, \citenamefont {van~der Wiel}, \citenamefont {Obata},\ and\ \citenamefont {Tarucha}}]{RN1760}%
  \BibitemOpen
  \bibfield  {author} {\bibinfo {author} {\bibfnamefont {Y.}~\bibnamefont {Tokura}}, \bibinfo {author} {\bibfnamefont {W.~G.}\ \bibnamefont {van~der Wiel}}, \bibinfo {author} {\bibfnamefont {T.}~\bibnamefont {Obata}},\ and\ \bibinfo {author} {\bibfnamefont {S.}~\bibnamefont {Tarucha}},\ }\bibfield  {title} {\enquote {\bibinfo {title} {Coherent single electron spin control in a slanting zeeman field},}\ }\href {https://doi.org/10.1103/PhysRevLett.96.047202} {\bibfield  {journal} {\bibinfo  {journal} {Phys. Rev. Lett.}\ }\textbf {\bibinfo {volume} {96}},\ \bibinfo {pages} {047202} (\bibinfo {year} {2006})}\BibitemShut {NoStop}%
\bibitem [{\citenamefont {Yoneda}\ \emph {et~al.}(2015)\citenamefont {Yoneda}, \citenamefont {Otsuka}, \citenamefont {Takakura}, \citenamefont {Pioro-LadriSre}, \citenamefont {Brunner}, \citenamefont {Lu}, \citenamefont {Nakajima}, \citenamefont {Obata}, \citenamefont {Noiri}, \citenamefont {Palmstrom}, \citenamefont {Gossard},\ and\ \citenamefont {Tarucha}}]{RN1578}%
  \BibitemOpen
  \bibfield  {author} {\bibinfo {author} {\bibfnamefont {J.}~\bibnamefont {Yoneda}}, \bibinfo {author} {\bibfnamefont {T.}~\bibnamefont {Otsuka}}, \bibinfo {author} {\bibfnamefont {T.}~\bibnamefont {Takakura}}, \bibinfo {author} {\bibfnamefont {M.}~\bibnamefont {Pioro-LadriSre}}, \bibinfo {author} {\bibfnamefont {R.}~\bibnamefont {Brunner}}, \bibinfo {author} {\bibfnamefont {H.}~\bibnamefont {Lu}}, \bibinfo {author} {\bibfnamefont {T.}~\bibnamefont {Nakajima}}, \bibinfo {author} {\bibfnamefont {T.}~\bibnamefont {Obata}}, \bibinfo {author} {\bibfnamefont {A.}~\bibnamefont {Noiri}}, \bibinfo {author} {\bibfnamefont {C.~J.}\ \bibnamefont {Palmstrom}}, \bibinfo {author} {\bibfnamefont {A.~C.}\ \bibnamefont {Gossard}},\ and\ \bibinfo {author} {\bibfnamefont {S.}~\bibnamefont {Tarucha}},\ }\bibfield  {title} {\enquote {\bibinfo {title} {Robust micromagnet design for fast electrical manipulations of single spins in quantum dots},}\ }\href {https://doi.org/10.7567/Apex.8.084401} {\bibfield  {journal} {\bibinfo  {journal} {Appl. Phys. Express}\ }\textbf {\bibinfo {volume} {8}},\ \bibinfo {pages} {084401} (\bibinfo {year} {2015})}\BibitemShut {NoStop}%
\bibitem [{\citenamefont {Yoneda}\ \emph {et~al.}(2021)\citenamefont {Yoneda}, \citenamefont {Huang}, \citenamefont {Feng}, \citenamefont {Yang}, \citenamefont {Chan}, \citenamefont {Tanttu}, \citenamefont {Gilbert}, \citenamefont {Leon}, \citenamefont {Hudson}, \citenamefont {Itoh}, \citenamefont {Morello}, \citenamefont {Bartlett}, \citenamefont {Laucht}, \citenamefont {Saraiva},\ and\ \citenamefont {Dzurak}}]{RN9241}%
  \BibitemOpen
  \bibfield  {author} {\bibinfo {author} {\bibfnamefont {J.}~\bibnamefont {Yoneda}}, \bibinfo {author} {\bibfnamefont {W.}~\bibnamefont {Huang}}, \bibinfo {author} {\bibfnamefont {M.}~\bibnamefont {Feng}}, \bibinfo {author} {\bibfnamefont {C.~H.}\ \bibnamefont {Yang}}, \bibinfo {author} {\bibfnamefont {K.~W.}\ \bibnamefont {Chan}}, \bibinfo {author} {\bibfnamefont {T.}~\bibnamefont {Tanttu}}, \bibinfo {author} {\bibfnamefont {W.}~\bibnamefont {Gilbert}}, \bibinfo {author} {\bibfnamefont {R.~C.~C.}\ \bibnamefont {Leon}}, \bibinfo {author} {\bibfnamefont {F.~E.}\ \bibnamefont {Hudson}}, \bibinfo {author} {\bibfnamefont {K.~M.}\ \bibnamefont {Itoh}}, \bibinfo {author} {\bibfnamefont {A.}~\bibnamefont {Morello}}, \bibinfo {author} {\bibfnamefont {S.~D.}\ \bibnamefont {Bartlett}}, \bibinfo {author} {\bibfnamefont {A.}~\bibnamefont {Laucht}}, \bibinfo {author} {\bibfnamefont {A.}~\bibnamefont {Saraiva}},\ and\ \bibinfo {author} {\bibfnamefont {A.~S.}\ \bibnamefont {Dzurak}},\ }\bibfield  {title} {\enquote {\bibinfo {title} {Coherent spin qubit transport in silicon},}\ }\href {https://doi.org/10.1038/s41467-021-24371-7} {\bibfield  {journal} {\bibinfo  {journal} {Nat. Commun.}\ }\textbf {\bibinfo {volume} {12}},\ \bibinfo {pages} {4114} (\bibinfo {year} {2021})}\BibitemShut {NoStop}%
\bibitem [{\citenamefont {De~Smet}\ \emph {et~al.}(2024)\citenamefont {De~Smet}, \citenamefont {Matsumoto}, \citenamefont {Zwerver}, \citenamefont {Tryputen}, \citenamefont {de~Snoo}, \citenamefont {Amitonov}, \citenamefont {Sammak}, \citenamefont {Samkharadze}, \citenamefont {G{\"u}l}, \citenamefont {Wasserman}, \citenamefont {Rimbach-Russ}, \citenamefont {Scappucci},\ and\ \citenamefont {Vandersypen}}]{RN12024}%
  \BibitemOpen
  \bibfield  {author} {\bibinfo {author} {\bibfnamefont {M.}~\bibnamefont {De~Smet}}, \bibinfo {author} {\bibfnamefont {Y.}~\bibnamefont {Matsumoto}}, \bibinfo {author} {\bibfnamefont {A.-M.~J.}\ \bibnamefont {Zwerver}}, \bibinfo {author} {\bibfnamefont {L.}~\bibnamefont {Tryputen}}, \bibinfo {author} {\bibfnamefont {S.~L.}\ \bibnamefont {de~Snoo}}, \bibinfo {author} {\bibfnamefont {S.~V.}\ \bibnamefont {Amitonov}}, \bibinfo {author} {\bibfnamefont {A.}~\bibnamefont {Sammak}}, \bibinfo {author} {\bibfnamefont {N.}~\bibnamefont {Samkharadze}}, \bibinfo {author} {\bibfnamefont {{\"O}.}~\bibnamefont {G{\"u}l}}, \bibinfo {author} {\bibfnamefont {R.~N.~M.}\ \bibnamefont {Wasserman}}, \bibinfo {author} {\bibfnamefont {M.}~\bibnamefont {Rimbach-Russ}}, \bibinfo {author} {\bibfnamefont {G.}~\bibnamefont {Scappucci}},\ and\ \bibinfo {author} {\bibfnamefont {L.~M.~K.}\ \bibnamefont {Vandersypen}},\ }\bibfield  {title} {\enquote {\bibinfo {title} {High-fidelity single-spin shuttling in silicon},}\ }\href@noop {} {\bibfield  {journal} {\bibinfo  {journal} {arXiv preprint arXiv:2406.07267}\ } (\bibinfo {year} {2024})}\BibitemShut {NoStop}%
\bibitem [{\citenamefont {van Riggelen-Doelman}\ \emph {et~al.}(2024)\citenamefont {van Riggelen-Doelman}, \citenamefont {Wang}, \citenamefont {de~Snoo}, \citenamefont {Lawrie}, \citenamefont {Hendrickx}, \citenamefont {Rimbach-Russ}, \citenamefont {Sammak}, \citenamefont {Scappucci}, \citenamefont {D{\'e}prez},\ and\ \citenamefont {Veldhorst}}]{RN12030}%
  \BibitemOpen
  \bibfield  {author} {\bibinfo {author} {\bibfnamefont {F.}~\bibnamefont {van Riggelen-Doelman}}, \bibinfo {author} {\bibfnamefont {C.-A.}\ \bibnamefont {Wang}}, \bibinfo {author} {\bibfnamefont {S.~L.}\ \bibnamefont {de~Snoo}}, \bibinfo {author} {\bibfnamefont {W.~I.~L.}\ \bibnamefont {Lawrie}}, \bibinfo {author} {\bibfnamefont {N.~W.}\ \bibnamefont {Hendrickx}}, \bibinfo {author} {\bibfnamefont {M.}~\bibnamefont {Rimbach-Russ}}, \bibinfo {author} {\bibfnamefont {A.}~\bibnamefont {Sammak}}, \bibinfo {author} {\bibfnamefont {G.}~\bibnamefont {Scappucci}}, \bibinfo {author} {\bibfnamefont {C.}~\bibnamefont {D{\'e}prez}},\ and\ \bibinfo {author} {\bibfnamefont {M.}~\bibnamefont {Veldhorst}},\ }\bibfield  {title} {\enquote {\bibinfo {title} {Coherent spin qubit shuttling through germanium quantum dots},}\ }\href {https://doi.org/10.1038/s41467-024-49358-y} {\bibfield  {journal} {\bibinfo  {journal} {Nat. Commun.}\ }\textbf {\bibinfo {volume} {15}},\ \bibinfo {pages} {5716} (\bibinfo {year} {2024})}\BibitemShut {NoStop}%
\bibitem [{\citenamefont {Yoneda}\ \emph {et~al.}(2018)\citenamefont {Yoneda}, \citenamefont {Takeda}, \citenamefont {Otsuka}, \citenamefont {Nakajima}, \citenamefont {Delbecq}, \citenamefont {Allison}, \citenamefont {Honda}, \citenamefont {Kodera}, \citenamefont {Oda}, \citenamefont {Hoshi}, \citenamefont {Usami}, \citenamefont {Itoh},\ and\ \citenamefont {Tarucha}}]{RN1540}%
  \BibitemOpen
  \bibfield  {author} {\bibinfo {author} {\bibfnamefont {J.}~\bibnamefont {Yoneda}}, \bibinfo {author} {\bibfnamefont {K.}~\bibnamefont {Takeda}}, \bibinfo {author} {\bibfnamefont {T.}~\bibnamefont {Otsuka}}, \bibinfo {author} {\bibfnamefont {T.}~\bibnamefont {Nakajima}}, \bibinfo {author} {\bibfnamefont {M.~R.}\ \bibnamefont {Delbecq}}, \bibinfo {author} {\bibfnamefont {G.}~\bibnamefont {Allison}}, \bibinfo {author} {\bibfnamefont {T.}~\bibnamefont {Honda}}, \bibinfo {author} {\bibfnamefont {T.}~\bibnamefont {Kodera}}, \bibinfo {author} {\bibfnamefont {S.}~\bibnamefont {Oda}}, \bibinfo {author} {\bibfnamefont {Y.}~\bibnamefont {Hoshi}}, \bibinfo {author} {\bibfnamefont {N.}~\bibnamefont {Usami}}, \bibinfo {author} {\bibfnamefont {K.~M.}\ \bibnamefont {Itoh}},\ and\ \bibinfo {author} {\bibfnamefont {S.}~\bibnamefont {Tarucha}},\ }\bibfield  {title} {\enquote {\bibinfo {title} {A quantum-dot spin qubit with coherence limited by charge noise and fidelity higher than 99.9$\%$},}\ }\href {https://doi.org/10.1038/s41565-017-0014-x} {\bibfield  {journal} {\bibinfo  {journal} {Nat. Nanotechnol.}\ }\textbf {\bibinfo {volume} {13}},\ \bibinfo {pages} {102--106} (\bibinfo {year} {2018})}\BibitemShut {NoStop}%
\bibitem [{\citenamefont {Knill}\ \emph {et~al.}(2008)\citenamefont {Knill}, \citenamefont {Leibfried}, \citenamefont {Reichle}, \citenamefont {Britton}, \citenamefont {Blakestad}, \citenamefont {Jost}, \citenamefont {Langer}, \citenamefont {Ozeri}, \citenamefont {Seidelin},\ and\ \citenamefont {Wineland}}]{RN2054}%
  \BibitemOpen
  \bibfield  {author} {\bibinfo {author} {\bibfnamefont {E.}~\bibnamefont {Knill}}, \bibinfo {author} {\bibfnamefont {D.}~\bibnamefont {Leibfried}}, \bibinfo {author} {\bibfnamefont {R.}~\bibnamefont {Reichle}}, \bibinfo {author} {\bibfnamefont {J.}~\bibnamefont {Britton}}, \bibinfo {author} {\bibfnamefont {R.~B.}\ \bibnamefont {Blakestad}}, \bibinfo {author} {\bibfnamefont {J.~D.}\ \bibnamefont {Jost}}, \bibinfo {author} {\bibfnamefont {C.}~\bibnamefont {Langer}}, \bibinfo {author} {\bibfnamefont {R.}~\bibnamefont {Ozeri}}, \bibinfo {author} {\bibfnamefont {S.}~\bibnamefont {Seidelin}},\ and\ \bibinfo {author} {\bibfnamefont {D.~J.}\ \bibnamefont {Wineland}},\ }\bibfield  {title} {\enquote {\bibinfo {title} {Randomized benchmarking of quantum gates},}\ }\href {https://doi.org/10.1103/PhysRevA.77.012307} {\bibfield  {journal} {\bibinfo  {journal} {Phys. Rev. A}\ }\textbf {\bibinfo {volume} {77}},\ \bibinfo {pages} {012307} (\bibinfo {year} {2008})}\BibitemShut {NoStop}%
\bibitem [{\citenamefont {Magesan}\ \emph {et~al.}(2012)\citenamefont {Magesan}, \citenamefont {Gambetta}, \citenamefont {Johnson}, \citenamefont {Ryan}, \citenamefont {Chow}, \citenamefont {Merkel}, \citenamefont {da~Silva}, \citenamefont {Keefe}, \citenamefont {Rothwell}, \citenamefont {Ohki}, \citenamefont {Ketchen},\ and\ \citenamefont {Steffen}}]{RN2042}%
  \BibitemOpen
  \bibfield  {author} {\bibinfo {author} {\bibfnamefont {E.}~\bibnamefont {Magesan}}, \bibinfo {author} {\bibfnamefont {J.~M.}\ \bibnamefont {Gambetta}}, \bibinfo {author} {\bibfnamefont {B.~R.}\ \bibnamefont {Johnson}}, \bibinfo {author} {\bibfnamefont {C.~A.}\ \bibnamefont {Ryan}}, \bibinfo {author} {\bibfnamefont {J.~M.}\ \bibnamefont {Chow}}, \bibinfo {author} {\bibfnamefont {S.~T.}\ \bibnamefont {Merkel}}, \bibinfo {author} {\bibfnamefont {M.~P.}\ \bibnamefont {da~Silva}}, \bibinfo {author} {\bibfnamefont {G.~A.}\ \bibnamefont {Keefe}}, \bibinfo {author} {\bibfnamefont {M.~B.}\ \bibnamefont {Rothwell}}, \bibinfo {author} {\bibfnamefont {T.~A.}\ \bibnamefont {Ohki}}, \bibinfo {author} {\bibfnamefont {M.~B.}\ \bibnamefont {Ketchen}},\ and\ \bibinfo {author} {\bibfnamefont {M.}~\bibnamefont {Steffen}},\ }\bibfield  {title} {\enquote {\bibinfo {title} {Efficient measurement of quantum gate error by interleaved randomized benchmarking},}\ }\href {https://doi.org/10.1103/PhysRevLett.109.080505} {\bibfield  {journal} {\bibinfo  {journal} {Phys. Rev. Lett.}\ }\textbf {\bibinfo {volume} {109}},\ \bibinfo {pages} {080505} (\bibinfo {year} {2012})}\BibitemShut {NoStop}%
\bibitem [{\citenamefont {Reed}\ \emph {et~al.}(2016)\citenamefont {Reed}, \citenamefont {Maune}, \citenamefont {Andrews}, \citenamefont {Borselli}, \citenamefont {Eng}, \citenamefont {Jura}, \citenamefont {Kiselev}, \citenamefont {Ladd}, \citenamefont {Merkel}, \citenamefont {Milosavljevic}, \citenamefont {Pritchett}, \citenamefont {Rakher}, \citenamefont {Ross}, \citenamefont {Schmitz}, \citenamefont {Smith}, \citenamefont {Wright}, \citenamefont {Gyure},\ and\ \citenamefont {Hunter}}]{RN6050}%
  \BibitemOpen
  \bibfield  {author} {\bibinfo {author} {\bibfnamefont {M.~D.}\ \bibnamefont {Reed}}, \bibinfo {author} {\bibfnamefont {B.~M.}\ \bibnamefont {Maune}}, \bibinfo {author} {\bibfnamefont {R.~W.}\ \bibnamefont {Andrews}}, \bibinfo {author} {\bibfnamefont {M.~G.}\ \bibnamefont {Borselli}}, \bibinfo {author} {\bibfnamefont {K.}~\bibnamefont {Eng}}, \bibinfo {author} {\bibfnamefont {M.~P.}\ \bibnamefont {Jura}}, \bibinfo {author} {\bibfnamefont {A.~A.}\ \bibnamefont {Kiselev}}, \bibinfo {author} {\bibfnamefont {T.~D.}\ \bibnamefont {Ladd}}, \bibinfo {author} {\bibfnamefont {S.~T.}\ \bibnamefont {Merkel}}, \bibinfo {author} {\bibfnamefont {I.}~\bibnamefont {Milosavljevic}}, \bibinfo {author} {\bibfnamefont {E.~J.}\ \bibnamefont {Pritchett}}, \bibinfo {author} {\bibfnamefont {M.~T.}\ \bibnamefont {Rakher}}, \bibinfo {author} {\bibfnamefont {R.~S.}\ \bibnamefont {Ross}}, \bibinfo {author} {\bibfnamefont {A.~E.}\ \bibnamefont {Schmitz}}, \bibinfo {author} {\bibfnamefont {A.}~\bibnamefont {Smith}}, \bibinfo {author} {\bibfnamefont {J.~A.}\ \bibnamefont {Wright}}, \bibinfo {author} {\bibfnamefont {M.~F.}\ \bibnamefont {Gyure}},\ and\ \bibinfo {author} {\bibfnamefont {A.~T.}\ \bibnamefont {Hunter}},\ }\bibfield  {title} {\enquote {\bibinfo {title} {Reduced sensitivity to charge noise in semiconductor spin qubits via symmetric operation},}\ }\href {https://doi.org/10.1103/PhysRevLett.116.110402} {\bibfield  {journal} {\bibinfo  {journal} {Phys. Rev. Lett.}\ }\textbf {\bibinfo {volume} {116}},\ \bibinfo {pages} {110402} (\bibinfo {year} {2016})}\BibitemShut {NoStop}%
\bibitem [{\citenamefont {Meunier}, \citenamefont {Calado},\ and\ \citenamefont {Vandersypen}(2011)}]{RN1481}%
  \BibitemOpen
  \bibfield  {author} {\bibinfo {author} {\bibfnamefont {T.}~\bibnamefont {Meunier}}, \bibinfo {author} {\bibfnamefont {V.~E.}\ \bibnamefont {Calado}},\ and\ \bibinfo {author} {\bibfnamefont {L.~M.~K.}\ \bibnamefont {Vandersypen}},\ }\bibfield  {title} {\enquote {\bibinfo {title} {Efficient controlled-phase gate for single-spin qubits in quantum dots},}\ }\href {https://doi.org/10.1103/PhysRevB.83.121403} {\bibfield  {journal} {\bibinfo  {journal} {Phys. Rev. B}\ }\textbf {\bibinfo {volume} {83}},\ \bibinfo {pages} {121403} (\bibinfo {year} {2011})}\BibitemShut {NoStop}%
\bibitem [{\citenamefont {Martins}\ \emph {et~al.}(2016)\citenamefont {Martins}, \citenamefont {Malinowski}, \citenamefont {Nissen}, \citenamefont {Barnes}, \citenamefont {Fallahi}, \citenamefont {Gardner}, \citenamefont {Manfra}, \citenamefont {Marcus},\ and\ \citenamefont {Kuemmeth}}]{RN9199}%
  \BibitemOpen
  \bibfield  {author} {\bibinfo {author} {\bibfnamefont {F.}~\bibnamefont {Martins}}, \bibinfo {author} {\bibfnamefont {F.~K.}\ \bibnamefont {Malinowski}}, \bibinfo {author} {\bibfnamefont {P.~D.}\ \bibnamefont {Nissen}}, \bibinfo {author} {\bibfnamefont {E.}~\bibnamefont {Barnes}}, \bibinfo {author} {\bibfnamefont {S.}~\bibnamefont {Fallahi}}, \bibinfo {author} {\bibfnamefont {G.~C.}\ \bibnamefont {Gardner}}, \bibinfo {author} {\bibfnamefont {M.~J.}\ \bibnamefont {Manfra}}, \bibinfo {author} {\bibfnamefont {C.~M.}\ \bibnamefont {Marcus}},\ and\ \bibinfo {author} {\bibfnamefont {F.}~\bibnamefont {Kuemmeth}},\ }\bibfield  {title} {\enquote {\bibinfo {title} {Noise suppression using symmetric exchange gates in spin qubits},}\ }\href {https://doi.org/10.1103/PhysRevLett.116.116801} {\bibfield  {journal} {\bibinfo  {journal} {Phys. Rev. Lett.}\ }\textbf {\bibinfo {volume} {116}},\ \bibinfo {pages} {116801} (\bibinfo {year} {2016})}\BibitemShut {NoStop}%
\bibitem [{\citenamefont {Takeda}\ \emph {et~al.}(2021)\citenamefont {Takeda}, \citenamefont {Noiri}, \citenamefont {Nakajima}, \citenamefont {Yoneda}, \citenamefont {Kobayashi},\ and\ \citenamefont {Tarucha}}]{RN9013}%
  \BibitemOpen
  \bibfield  {author} {\bibinfo {author} {\bibfnamefont {K.}~\bibnamefont {Takeda}}, \bibinfo {author} {\bibfnamefont {A.}~\bibnamefont {Noiri}}, \bibinfo {author} {\bibfnamefont {T.}~\bibnamefont {Nakajima}}, \bibinfo {author} {\bibfnamefont {J.}~\bibnamefont {Yoneda}}, \bibinfo {author} {\bibfnamefont {T.}~\bibnamefont {Kobayashi}},\ and\ \bibinfo {author} {\bibfnamefont {S.}~\bibnamefont {Tarucha}},\ }\bibfield  {title} {\enquote {\bibinfo {title} {Quantum tomography of an entangled three-qubit state in silicon},}\ }\href {https://doi.org/10.1038/s41565-021-00925-0} {\bibfield  {journal} {\bibinfo  {journal} {Nat. Nanotechnol.}\ }\textbf {\bibinfo {volume} {16}},\ \bibinfo {pages} {965--969} (\bibinfo {year} {2021})}\BibitemShut {NoStop}%
\bibitem [{\citenamefont {Takeda}\ \emph {et~al.}(2022)\citenamefont {Takeda}, \citenamefont {Noiri}, \citenamefont {Nakajima}, \citenamefont {Kobayashi},\ and\ \citenamefont {Tarucha}}]{RN11423}%
  \BibitemOpen
  \bibfield  {author} {\bibinfo {author} {\bibfnamefont {K.}~\bibnamefont {Takeda}}, \bibinfo {author} {\bibfnamefont {A.}~\bibnamefont {Noiri}}, \bibinfo {author} {\bibfnamefont {T.}~\bibnamefont {Nakajima}}, \bibinfo {author} {\bibfnamefont {T.}~\bibnamefont {Kobayashi}},\ and\ \bibinfo {author} {\bibfnamefont {S.}~\bibnamefont {Tarucha}},\ }\bibfield  {title} {\enquote {\bibinfo {title} {Quantum error correction with silicon spin qubits},}\ }\href {https://doi.org/10.1038/s41586-022-04986-6} {\bibfield  {journal} {\bibinfo  {journal} {Nature}\ }\textbf {\bibinfo {volume} {608}},\ \bibinfo {pages} {682--686} (\bibinfo {year} {2022})}\BibitemShut {NoStop}%
\bibitem [{\citenamefont {Gaebler}\ \emph {et~al.}(2012)\citenamefont {Gaebler}, \citenamefont {Meier}, \citenamefont {Tan}, \citenamefont {Bowler}, \citenamefont {Lin}, \citenamefont {Hanneke}, \citenamefont {Jost}, \citenamefont {Home}, \citenamefont {Knill}, \citenamefont {Leibfried},\ and\ \citenamefont {Wineland}}]{RN2045}%
  \BibitemOpen
  \bibfield  {author} {\bibinfo {author} {\bibfnamefont {J.~P.}\ \bibnamefont {Gaebler}}, \bibinfo {author} {\bibfnamefont {A.~M.}\ \bibnamefont {Meier}}, \bibinfo {author} {\bibfnamefont {T.~R.}\ \bibnamefont {Tan}}, \bibinfo {author} {\bibfnamefont {R.}~\bibnamefont {Bowler}}, \bibinfo {author} {\bibfnamefont {Y.}~\bibnamefont {Lin}}, \bibinfo {author} {\bibfnamefont {D.}~\bibnamefont {Hanneke}}, \bibinfo {author} {\bibfnamefont {J.~D.}\ \bibnamefont {Jost}}, \bibinfo {author} {\bibfnamefont {J.~P.}\ \bibnamefont {Home}}, \bibinfo {author} {\bibfnamefont {E.}~\bibnamefont {Knill}}, \bibinfo {author} {\bibfnamefont {D.}~\bibnamefont {Leibfried}},\ and\ \bibinfo {author} {\bibfnamefont {D.~J.}\ \bibnamefont {Wineland}},\ }\bibfield  {title} {\enquote {\bibinfo {title} {Randomized benchmarking of multiqubit gates},}\ }\href {https://doi.org/10.1103/PhysRevLett.108.260503} {\bibfield  {journal} {\bibinfo  {journal} {Phys. Rev. Lett.}\ }\textbf {\bibinfo {volume} {108}},\ \bibinfo {pages} {260503} (\bibinfo {year} {2012})}\BibitemShut {NoStop}%
\bibitem [{\citenamefont {Barends}\ \emph {et~al.}(2014)\citenamefont {Barends}, \citenamefont {Kelly}, \citenamefont {Megrant}, \citenamefont {Veitia}, \citenamefont {Sank}, \citenamefont {Jeffrey}, \citenamefont {White}, \citenamefont {Mutus}, \citenamefont {Fowler}, \citenamefont {Campbell}, \citenamefont {Chen}, \citenamefont {Chen}, \citenamefont {Chiaro}, \citenamefont {Dunsworth}, \citenamefont {Neill}, \citenamefont {O'Malley}, \citenamefont {Roushan}, \citenamefont {Vainsencher}, \citenamefont {Wenner}, \citenamefont {Korotkov}, \citenamefont {Cleland},\ and\ \citenamefont {Martinis}}]{RN11996}%
  \BibitemOpen
  \bibfield  {author} {\bibinfo {author} {\bibfnamefont {R.}~\bibnamefont {Barends}}, \bibinfo {author} {\bibfnamefont {J.}~\bibnamefont {Kelly}}, \bibinfo {author} {\bibfnamefont {A.}~\bibnamefont {Megrant}}, \bibinfo {author} {\bibfnamefont {A.}~\bibnamefont {Veitia}}, \bibinfo {author} {\bibfnamefont {D.}~\bibnamefont {Sank}}, \bibinfo {author} {\bibfnamefont {E.}~\bibnamefont {Jeffrey}}, \bibinfo {author} {\bibfnamefont {T.~C.}\ \bibnamefont {White}}, \bibinfo {author} {\bibfnamefont {J.}~\bibnamefont {Mutus}}, \bibinfo {author} {\bibfnamefont {A.~G.}\ \bibnamefont {Fowler}}, \bibinfo {author} {\bibfnamefont {B.}~\bibnamefont {Campbell}}, \bibinfo {author} {\bibfnamefont {Y.}~\bibnamefont {Chen}}, \bibinfo {author} {\bibfnamefont {Z.}~\bibnamefont {Chen}}, \bibinfo {author} {\bibfnamefont {B.}~\bibnamefont {Chiaro}}, \bibinfo {author} {\bibfnamefont {A.}~\bibnamefont {Dunsworth}}, \bibinfo {author} {\bibfnamefont {C.}~\bibnamefont {Neill}}, \bibinfo {author} {\bibfnamefont {P.}~\bibnamefont {O'Malley}}, \bibinfo {author} {\bibfnamefont {P.}~\bibnamefont {Roushan}}, \bibinfo {author} {\bibfnamefont {A.}~\bibnamefont {Vainsencher}}, \bibinfo {author} {\bibfnamefont {J.}~\bibnamefont {Wenner}}, \bibinfo {author} {\bibfnamefont {A.~N.}\ \bibnamefont {Korotkov}}, \bibinfo {author} {\bibfnamefont {A.~N.}\ \bibnamefont {Cleland}},\ and\ \bibinfo {author} {\bibfnamefont {J.~M.}\ \bibnamefont {Martinis}},\ }\bibfield  {title} {\enquote {\bibinfo {title} {Superconducting quantum circuits at the surface code threshold for fault tolerance},}\ }\href {https://doi.org/10.1038/nature13171} {\bibfield  {journal} {\bibinfo  {journal} {Nature}\ }\textbf {\bibinfo {volume} {508}},\ \bibinfo {pages} {500--503} (\bibinfo {year} {2014})}\BibitemShut {NoStop}%
\bibitem [{\citenamefont {Hendrickx}\ \emph {et~al.}(2021)\citenamefont {Hendrickx}, \citenamefont {Lawrie}, \citenamefont {Russ}, \citenamefont {van Riggelen}, \citenamefont {de~Snoo}, \citenamefont {Schouten}, \citenamefont {Sammak}, \citenamefont {Scappucci},\ and\ \citenamefont {Veldhorst}}]{RN8962}%
  \BibitemOpen
  \bibfield  {author} {\bibinfo {author} {\bibfnamefont {N.~W.}\ \bibnamefont {Hendrickx}}, \bibinfo {author} {\bibfnamefont {W.~I.~L.}\ \bibnamefont {Lawrie}}, \bibinfo {author} {\bibfnamefont {M.}~\bibnamefont {Russ}}, \bibinfo {author} {\bibfnamefont {F.}~\bibnamefont {van Riggelen}}, \bibinfo {author} {\bibfnamefont {S.~L.}\ \bibnamefont {de~Snoo}}, \bibinfo {author} {\bibfnamefont {R.~N.}\ \bibnamefont {Schouten}}, \bibinfo {author} {\bibfnamefont {A.}~\bibnamefont {Sammak}}, \bibinfo {author} {\bibfnamefont {G.}~\bibnamefont {Scappucci}},\ and\ \bibinfo {author} {\bibfnamefont {M.}~\bibnamefont {Veldhorst}},\ }\bibfield  {title} {\enquote {\bibinfo {title} {A four-qubit germanium quantum processor},}\ }\href {https://doi.org/10.1038/s41586-021-03332-6} {\bibfield  {journal} {\bibinfo  {journal} {Nature}\ }\textbf {\bibinfo {volume} {591}},\ \bibinfo {pages} {580--585} (\bibinfo {year} {2021})}\BibitemShut {NoStop}%
\bibitem [{\citenamefont {Noiri}\ \emph {et~al.}(2022{\natexlab{b}})\citenamefont {Noiri}, \citenamefont {Takeda}, \citenamefont {Nakajima}, \citenamefont {Kobayashi}, \citenamefont {Sammak}, \citenamefont {Scappucci},\ and\ \citenamefont {Tarucha}}]{RN11494}%
  \BibitemOpen
  \bibfield  {author} {\bibinfo {author} {\bibfnamefont {A.}~\bibnamefont {Noiri}}, \bibinfo {author} {\bibfnamefont {K.}~\bibnamefont {Takeda}}, \bibinfo {author} {\bibfnamefont {T.}~\bibnamefont {Nakajima}}, \bibinfo {author} {\bibfnamefont {T.}~\bibnamefont {Kobayashi}}, \bibinfo {author} {\bibfnamefont {A.}~\bibnamefont {Sammak}}, \bibinfo {author} {\bibfnamefont {G.}~\bibnamefont {Scappucci}},\ and\ \bibinfo {author} {\bibfnamefont {S.}~\bibnamefont {Tarucha}},\ }\bibfield  {title} {\enquote {\bibinfo {title} {A shuttling-based two-qubit logic gate for linking distant silicon quantum processors},}\ }\href {https://doi.org/10.1038/s41467-022-33453-z} {\bibfield  {journal} {\bibinfo  {journal} {Nat. Commun.}\ }\textbf {\bibinfo {volume} {13}},\ \bibinfo {pages} {5740} (\bibinfo {year} {2022}{\natexlab{b}})}\BibitemShut {NoStop}%
\bibitem [{\citenamefont {Huang}\ \emph {et~al.}(2019)\citenamefont {Huang}, \citenamefont {Yang}, \citenamefont {Chan}, \citenamefont {Tanttu}, \citenamefont {Hensen}, \citenamefont {Leon}, \citenamefont {Fogarty}, \citenamefont {Hwang}, \citenamefont {Hudson}, \citenamefont {Itoh}, \citenamefont {Morello}, \citenamefont {Laucht},\ and\ \citenamefont {Dzurak}}]{RN1205}%
  \BibitemOpen
  \bibfield  {author} {\bibinfo {author} {\bibfnamefont {W.}~\bibnamefont {Huang}}, \bibinfo {author} {\bibfnamefont {C.~H.}\ \bibnamefont {Yang}}, \bibinfo {author} {\bibfnamefont {K.~W.}\ \bibnamefont {Chan}}, \bibinfo {author} {\bibfnamefont {T.}~\bibnamefont {Tanttu}}, \bibinfo {author} {\bibfnamefont {B.}~\bibnamefont {Hensen}}, \bibinfo {author} {\bibfnamefont {R.~C.~C.}\ \bibnamefont {Leon}}, \bibinfo {author} {\bibfnamefont {M.~A.}\ \bibnamefont {Fogarty}}, \bibinfo {author} {\bibfnamefont {J.~C.~C.}\ \bibnamefont {Hwang}}, \bibinfo {author} {\bibfnamefont {F.~E.}\ \bibnamefont {Hudson}}, \bibinfo {author} {\bibfnamefont {K.~M.}\ \bibnamefont {Itoh}}, \bibinfo {author} {\bibfnamefont {A.}~\bibnamefont {Morello}}, \bibinfo {author} {\bibfnamefont {A.}~\bibnamefont {Laucht}},\ and\ \bibinfo {author} {\bibfnamefont {A.~S.}\ \bibnamefont {Dzurak}},\ }\bibfield  {title} {\enquote {\bibinfo {title} {Fidelity benchmarks for two-qubit gates in silicon},}\ }\href {https://doi.org/10.1038/s41586-019-1197-0} {\bibfield  {journal} {\bibinfo  {journal} {Nature}\ }\textbf {\bibinfo {volume} {569}},\ \bibinfo {pages} {532--536} (\bibinfo {year} {2019})}\BibitemShut {NoStop}%
\bibitem [{\citenamefont {Aldeghi}, \citenamefont {Allenspach},\ and\ \citenamefont {Salis}(2023)}]{RN12049}%
  \BibitemOpen
  \bibfield  {author} {\bibinfo {author} {\bibfnamefont {M.}~\bibnamefont {Aldeghi}}, \bibinfo {author} {\bibfnamefont {R.}~\bibnamefont {Allenspach}},\ and\ \bibinfo {author} {\bibfnamefont {G.}~\bibnamefont {Salis}},\ }\bibfield  {title} {\enquote {\bibinfo {title} {Modular nanomagnet design for spin qubits confined in a linear chain},}\ }\href {https://doi.org/10.1063/5.0139670} {\bibfield  {journal} {\bibinfo  {journal} {Appl. Phys. Lett.}\ }\textbf {\bibinfo {volume} {122}},\ \bibinfo {pages} {134003} (\bibinfo {year} {2023})}\BibitemShut {NoStop}%
\bibitem [{\citenamefont {Dumoulin~Stuyck}\ \emph {et~al.}(2021)\citenamefont {Dumoulin~Stuyck}, \citenamefont {Mohiyaddin}, \citenamefont {Li}, \citenamefont {Heyns}, \citenamefont {Govoreanu},\ and\ \citenamefont {Radu}}]{RN10499}%
  \BibitemOpen
  \bibfield  {author} {\bibinfo {author} {\bibfnamefont {N.~I.}\ \bibnamefont {Dumoulin~Stuyck}}, \bibinfo {author} {\bibfnamefont {F.~A.}\ \bibnamefont {Mohiyaddin}}, \bibinfo {author} {\bibfnamefont {R.}~\bibnamefont {Li}}, \bibinfo {author} {\bibfnamefont {M.}~\bibnamefont {Heyns}}, \bibinfo {author} {\bibfnamefont {B.}~\bibnamefont {Govoreanu}},\ and\ \bibinfo {author} {\bibfnamefont {I.~P.}\ \bibnamefont {Radu}},\ }\bibfield  {title} {\enquote {\bibinfo {title} {Low dephasing and robust micromagnet designs for silicon spin qubits},}\ }\href {https://doi.org/10.1063/5.0059939} {\bibfield  {journal} {\bibinfo  {journal} {Appl. Phys. Lett.}\ }\textbf {\bibinfo {volume} {119}},\ \bibinfo {pages} {094001} (\bibinfo {year} {2021})}\BibitemShut {NoStop}%
\bibitem [{\citenamefont {Shulman}\ \emph {et~al.}(2014)\citenamefont {Shulman}, \citenamefont {Harvey}, \citenamefont {Nichol}, \citenamefont {Bartlett}, \citenamefont {Doherty}, \citenamefont {Umansky},\ and\ \citenamefont {Yacoby}}]{RN6097}%
  \BibitemOpen
  \bibfield  {author} {\bibinfo {author} {\bibfnamefont {M.~D.}\ \bibnamefont {Shulman}}, \bibinfo {author} {\bibfnamefont {S.~P.}\ \bibnamefont {Harvey}}, \bibinfo {author} {\bibfnamefont {J.~M.}\ \bibnamefont {Nichol}}, \bibinfo {author} {\bibfnamefont {S.~D.}\ \bibnamefont {Bartlett}}, \bibinfo {author} {\bibfnamefont {A.~C.}\ \bibnamefont {Doherty}}, \bibinfo {author} {\bibfnamefont {V.}~\bibnamefont {Umansky}},\ and\ \bibinfo {author} {\bibfnamefont {A.}~\bibnamefont {Yacoby}},\ }\bibfield  {title} {\enquote {\bibinfo {title} {Suppressing qubit dephasing using real-time hamiltonian estimation},}\ }\href {https://doi.org/10.1038/ncomms6156} {\bibfield  {journal} {\bibinfo  {journal} {Nat. Commun.}\ }\textbf {\bibinfo {volume} {5}},\ \bibinfo {pages} {5156} (\bibinfo {year} {2014})}\BibitemShut {NoStop}%
\bibitem [{\citenamefont {Nakajima}\ \emph {et~al.}(2020)\citenamefont {Nakajima}, \citenamefont {Noiri}, \citenamefont {Kawasaki}, \citenamefont {Yoneda}, \citenamefont {Stano}, \citenamefont {Amaha}, \citenamefont {Otsuka}, \citenamefont {Takeda}, \citenamefont {Delbecq}, \citenamefont {Allison}, \citenamefont {Ludwig}, \citenamefont {Wieck}, \citenamefont {Loss},\ and\ \citenamefont {Tarucha}}]{RN6006}%
  \BibitemOpen
  \bibfield  {author} {\bibinfo {author} {\bibfnamefont {T.}~\bibnamefont {Nakajima}}, \bibinfo {author} {\bibfnamefont {A.}~\bibnamefont {Noiri}}, \bibinfo {author} {\bibfnamefont {K.}~\bibnamefont {Kawasaki}}, \bibinfo {author} {\bibfnamefont {J.}~\bibnamefont {Yoneda}}, \bibinfo {author} {\bibfnamefont {P.}~\bibnamefont {Stano}}, \bibinfo {author} {\bibfnamefont {S.}~\bibnamefont {Amaha}}, \bibinfo {author} {\bibfnamefont {T.}~\bibnamefont {Otsuka}}, \bibinfo {author} {\bibfnamefont {K.}~\bibnamefont {Takeda}}, \bibinfo {author} {\bibfnamefont {M.~R.}\ \bibnamefont {Delbecq}}, \bibinfo {author} {\bibfnamefont {G.}~\bibnamefont {Allison}}, \bibinfo {author} {\bibfnamefont {A.}~\bibnamefont {Ludwig}}, \bibinfo {author} {\bibfnamefont {A.~D.}\ \bibnamefont {Wieck}}, \bibinfo {author} {\bibfnamefont {D.}~\bibnamefont {Loss}},\ and\ \bibinfo {author} {\bibfnamefont {S.}~\bibnamefont {Tarucha}},\ }\bibfield  {title} {\enquote {\bibinfo {title} {Coherence of a driven electron spin qubit actively decoupled from quasistatic noise},}\ }\href {https://doi.org/10.1103/PhysRevX.10.011060} {\bibfield  {journal} {\bibinfo  {journal} {Phys. Rev. X}\ }\textbf {\bibinfo {volume} {10}},\ \bibinfo {pages} {011060} (\bibinfo {year} {2020})}\BibitemShut {NoStop}%
\bibitem [{\citenamefont {Kim}\ \emph {et~al.}(2022)\citenamefont {Kim}, \citenamefont {Yun}, \citenamefont {Jang}, \citenamefont {Jang}, \citenamefont {Park}, \citenamefont {Song}, \citenamefont {Cho}, \citenamefont {Sim}, \citenamefont {Sohn}, \citenamefont {Jung}, \citenamefont {Umansky},\ and\ \citenamefont {Kim}}]{RN12039}%
  \BibitemOpen
  \bibfield  {author} {\bibinfo {author} {\bibfnamefont {J.}~\bibnamefont {Kim}}, \bibinfo {author} {\bibfnamefont {J.}~\bibnamefont {Yun}}, \bibinfo {author} {\bibfnamefont {W.}~\bibnamefont {Jang}}, \bibinfo {author} {\bibfnamefont {H.}~\bibnamefont {Jang}}, \bibinfo {author} {\bibfnamefont {J.}~\bibnamefont {Park}}, \bibinfo {author} {\bibfnamefont {Y.}~\bibnamefont {Song}}, \bibinfo {author} {\bibfnamefont {M.~K.}\ \bibnamefont {Cho}}, \bibinfo {author} {\bibfnamefont {S.}~\bibnamefont {Sim}}, \bibinfo {author} {\bibfnamefont {H.}~\bibnamefont {Sohn}}, \bibinfo {author} {\bibfnamefont {H.}~\bibnamefont {Jung}}, \bibinfo {author} {\bibfnamefont {V.}~\bibnamefont {Umansky}},\ and\ \bibinfo {author} {\bibfnamefont {D.}~\bibnamefont {Kim}},\ }\bibfield  {title} {\enquote {\bibinfo {title} {Approaching ideal visibility in singlet-triplet qubit operations using energy-selective tunneling-based hamiltonian estimation},}\ }\href {https://doi.org/10.1103/PhysRevLett.129.040501} {\bibfield  {journal} {\bibinfo  {journal} {Phys. Rev. Lett.}\ }\textbf {\bibinfo {volume} {129}},\ \bibinfo {pages} {040501} (\bibinfo {year} {2022})}\BibitemShut {NoStop}%
\bibitem [{\citenamefont {Yun}\ \emph {et~al.}(2023)\citenamefont {Yun}, \citenamefont {Park}, \citenamefont {Jang}, \citenamefont {Kim}, \citenamefont {Jang}, \citenamefont {Song}, \citenamefont {Cho}, \citenamefont {Sohn}, \citenamefont {Jung}, \citenamefont {Umansky},\ and\ \citenamefont {Kim}}]{RN11685}%
  \BibitemOpen
  \bibfield  {author} {\bibinfo {author} {\bibfnamefont {J.}~\bibnamefont {Yun}}, \bibinfo {author} {\bibfnamefont {J.}~\bibnamefont {Park}}, \bibinfo {author} {\bibfnamefont {H.}~\bibnamefont {Jang}}, \bibinfo {author} {\bibfnamefont {J.}~\bibnamefont {Kim}}, \bibinfo {author} {\bibfnamefont {W.}~\bibnamefont {Jang}}, \bibinfo {author} {\bibfnamefont {Y.}~\bibnamefont {Song}}, \bibinfo {author} {\bibfnamefont {M.-K.}\ \bibnamefont {Cho}}, \bibinfo {author} {\bibfnamefont {H.}~\bibnamefont {Sohn}}, \bibinfo {author} {\bibfnamefont {H.}~\bibnamefont {Jung}}, \bibinfo {author} {\bibfnamefont {V.}~\bibnamefont {Umansky}},\ and\ \bibinfo {author} {\bibfnamefont {D.}~\bibnamefont {Kim}},\ }\bibfield  {title} {\enquote {\bibinfo {title} {Probing two-qubit capacitive interactions beyond bilinear regime using dual hamiltonian parameter estimations},}\ }\href {https://doi.org/10.1038/s41534-023-00699-4} {\bibfield  {journal} {\bibinfo  {journal} {npj Quantum Inf.}\ }\textbf {\bibinfo {volume} {9}},\ \bibinfo {pages} {30} (\bibinfo {year} {2023})}\BibitemShut {NoStop}%
\bibitem [{\citenamefont {Noiri}\ \emph {et~al.}(2020)\citenamefont {Noiri}, \citenamefont {Takeda}, \citenamefont {Yoneda}, \citenamefont {Nakajima}, \citenamefont {Kodera},\ and\ \citenamefont {Tarucha}}]{RN8623}%
  \BibitemOpen
  \bibfield  {author} {\bibinfo {author} {\bibfnamefont {A.}~\bibnamefont {Noiri}}, \bibinfo {author} {\bibfnamefont {K.}~\bibnamefont {Takeda}}, \bibinfo {author} {\bibfnamefont {J.}~\bibnamefont {Yoneda}}, \bibinfo {author} {\bibfnamefont {T.}~\bibnamefont {Nakajima}}, \bibinfo {author} {\bibfnamefont {T.}~\bibnamefont {Kodera}},\ and\ \bibinfo {author} {\bibfnamefont {S.}~\bibnamefont {Tarucha}},\ }\bibfield  {title} {\enquote {\bibinfo {title} {Radio-frequency-detected fast charge sensing in undoped silicon quantum dots},}\ }\href {https://doi.org/10.1021/acs.nanolett.9b03847} {\bibfield  {journal} {\bibinfo  {journal} {Nano Lett.}\ }\textbf {\bibinfo {volume} {20}},\ \bibinfo {pages} {947--952} (\bibinfo {year} {2020})}\BibitemShut {NoStop}%
\bibitem [{\citenamefont {Connors}, \citenamefont {Nelson},\ and\ \citenamefont {Nichol}(2020)}]{RN8062}%
  \BibitemOpen
  \bibfield  {author} {\bibinfo {author} {\bibfnamefont {E.~J.}\ \bibnamefont {Connors}}, \bibinfo {author} {\bibfnamefont {J.~J.}\ \bibnamefont {Nelson}},\ and\ \bibinfo {author} {\bibfnamefont {J.~M.}\ \bibnamefont {Nichol}},\ }\bibfield  {title} {\enquote {\bibinfo {title} {Rapid high-fidelity spin-state readout in $\text{Si/SiGe}$ quantum dots via rf reflectometry},}\ }\href {https://doi.org/10.1103/PhysRevApplied.13.024019} {\bibfield  {journal} {\bibinfo  {journal} {Phys. Rev. Appl.}\ }\textbf {\bibinfo {volume} {13}},\ \bibinfo {pages} {024019} (\bibinfo {year} {2020})}\BibitemShut {NoStop}%
\end{thebibliography}%
\end{document}